\documentclass[10pt,journal,compsoc]{IEEEtran}
\ifCLASSOPTIONcompsoc
  \usepackage[nocompress]{cite}
\else
  \usepackage{cite}
\fi
\ifCLASSINFOpdf
  \usepackage[pdftex]{graphicx}
  \graphicspath{{../pdf/}{../jpeg/}}
  \DeclareGraphicsExtensions{.pdf,.jpeg,.png}
\else
\fi
\usepackage{colortbl,booktabs} 
\usepackage{multirow}
\usepackage[justification=centering]{caption} 
\usepackage[numbers]{natbib}
\usepackage[resetlabels]{multibib}
\usepackage{subfigure}
\newcites{sr}{Reference}
\usepackage[framemethod=TikZ]{mdframed}

\begin{document}

\title{Metrics for Software Process Simulation Modeling}

\author{Bohan Liu,
        He Zhang,
        Liming Dong,
        Zhiqi Wang,
        Shanshan Li
\IEEEcompsocitemizethanks{\IEEEcompsocthanksitem
B. Liu, H. Zhang, L. Dong, Z. Wang, and S. Li are with State Key Laboratory of Novel Software Technology, Software Institute, Nanjing University, Nanjing, Jiangsu, China. \protect\\
E-mail: bohanliu@nju.edu.cn, hezhang@nju.edu.cn, lmdongmg@gmail.com, 502022320013@smail.nju.edu.cn, lss@nju.edu.cn}}

%
%

\markboth{IEEE Transactions on Software Engineering, October~2018}%
{Shell \MakeLowercase{\textit{et al.}}: IEEE Transactions on Software Engineering}
%



\IEEEtitleabstractindextext{%
\begin{abstract}
\textbf{\emph{Background}}: Software Process Simulation (SPS) has become an effective tool for software process management and improvement. However, its adoption in industry is less than what the research community expected due to the burden of measurement cost and the high demand for domain knowledge. The difficulty of extracting appropriate metrics with real data from process enactment is one of the great challenges.

\textbf{\emph{Objective}}: We aim to provide evidence-based support of the process metrics for software process (simulation) modeling. 

\textbf{\emph{Method}}: A systematic literature review was performed by extending our previous review series to draw a comprehensive understanding of the metrics for process modeling following a meta-model of ontology of metrics in SPS.

\textbf{\emph{Results}}: We identified 145 process modeling studies that collectively involve 2130 metrics and classified them using the coding technique. Two diagrams which illustrate the high frequency causal relationships used between metrics are proposed in terms of two hierarchical levels of modeling purposes. The specific metrics of different paradigms are compared, and the main difference is that Discrete-Event Simulation (DES) and Agent-Based Simulation (ABS) can provide more detailed simulations from the perspective of development activities and individual developers whilst System Dynamics (SD) tends to use the mean value as an alternative. We revisited the data issues encountered in SPS data preparing phases, as well as identified the corresponding strategies.

\textbf{\emph{Conclusion}}: The results of this study provide process modelers with an evidence-based reference of the identification and the use of metrics in SPS modeling, and further contribute to the development of the body of knowledge on software metrics in the context of process modeling. Furthermore, this study is not limited to process simulation but can be extended to software process modeling, in general. Taking simulation metrics as standards and references can further motivate and guide software developers to improve the collection, governance, and application of process data in practice.
\end{abstract}

}

\maketitle

\IEEEdisplaynontitleabstractindextext

%
\IEEEpeerreviewmaketitle


%
%
%
%

 

\section{Introduction}

\IEEEPARstart{S}{oftware} process models are built to gain insights into software processes so that we can predict, modify or control them~\cite{Kellner1999Software}. Software process model can be either a static (descriptive) model or a dynamic (simulation) model whose behavior changes over time. The simulation requires more information and knowledge, but it can simulate the real world in more detail. It has been widely claimed and accepted that SPS is an effective tool in support of software process management and improvement. Since Abdel-Hamid and Madnick~\cite{Abdel1991Software} introduced Software Process Simulation (SPS) to Software Engineering (SE) in the 1980s, there have been a large number of studies published in the community, including quite a few industrial cases. Ahmed et al.~\cite{Ahmed2008Software} conducted a survey to investigate the state-of-practice of simulation practice in SPS, nearly half of the respondents (8$/$17) are from industry. Furthermore, researchers have applied the SPS technique within the integration of the capacity maturity model (CMMI) for process optimization~\cite{Birkh2010An,Crespo2012Decision,Li2012An}, the SPS was recognized as the key to achieving levels 4 and 5~\cite{Raffo1999Software}. In addition to CMMI, Mishra et al.~\cite{MishraM16} built a system dynamics model to understand the global software development of the Indian software industry. The system dynamics technology is also applied to choose the best gate timing strategy in new product development projects~\citesr{Van2017Measuring}\footnotemark[1]\footnotetext[1]{We use a distinct citation format to distinguish the reviewed studies from other references.}.

Zhang et.al~\cite{Zhang2011Impact} highlight benefits of SPS for various purposes such as prediction, process investigation, technology evaluation, and risk management. 
To achieve the modeling purposes, an SPS model may involve a number of (sometimes even hundreds of) metrics. The identification of the metrics and their relationships needed in a specific process model is a challenging task, particularly for novice modelers, and the collection of the quality data on these metrics is even effort-consuming. 

The panel of domain experts in SPS indicates that a prerequisite for building SPS models in the industry is that companies are able to analyze their information needs~\cite{ZhangCrossroads14}. It accounts for most of the cost to identify the appropriate metrics and gather the corresponding data. The analysis of information, as well as the identification of metrics, requires considerable knowledge and skills. Hence, the software process community encourages the development of the knowledge base and the model library with the common set of process metrics as the key component to unleash reusability~\cite{ZhangCrossroads14}. There are a number of SPS studies indicated the problems on metrics, some of the evidence are presented as follows:


\emph{``For some of the relevant data it is hardly possible to determine the necessary information in real-life projects ... we are elaborating approaches to take such human attributes into account, which are not directly observable, and to consider them in the quantitative logic of the model.''}~\citesr{Neu2003Creating}.

\emph{``Obtaining the quantitative data is another difficulty regardless of developing the simulation model.''}~\citesr{Park2007Deriving}

The challenging task on modeling metrics can be twofold: 1) identifying key metrics from real-process based on the domain knowledge and the data available; 2) collecting and mining the required data for measurement. Knowing what metrics were used in SPS modeling is a prerequisite for studying these two challenges. To the best of our knowledge, no secondary study that investigates metrics is dedicated to SPS modeling yet, although many papers and books~\cite{Gomez2006,Portobellini2008,Kitchenham2010WUS,Abilio2012A,Kupiainen2015,Nu2017Source,Meidan18} have been published on the topic of software metrics over the past decades. It motivated us to create an evidence-based view of the metrics adopted in SPS models to contribute to the development of the body of knowledge on software metrics in the context of process modeling. 

Therefore, the objective of this study and its follow-ups are to relieve the high burden and cost of SPS modeling by systematically identifying the modeling metrics of the exemplar process models and the experiences extracted from the relevant literature available. Although this research takes the metrics in SPS modeling as the research object, it is also applicable to general software process modeling as well as general software process measurement. The process of building a simulation model usually consists of the following steps. In the process of building an executable simulation model, a static model is often an essential intermediate product. From an evaluation point of view, we require that the descriptive model is semantically correct, while we need to assess the appropriateness and fidelity of the simulation model, as its output is a distribution of values for a specific project~\cite{Lindland1994Understanding,KitchenhamPLJ05}. From a modeling perspective, the building of descriptive model is to abstract, collect and transform fragments of the real world based on the incomplete knowledge we have gained so far~\cite{KUIPERS1989Qualitative}. Simulation models require detailed understanding of the processes they simulate, as well as the reliable data for their initial construction. For example, a set of variables needs to be specified that represents a continuously differentiable function of time~\citesr{ZhangHKJ06,Zhang2009Qualitative}. Hence, SPS modeling associates with a higher standard of metric than static process modeling. The metrics used in SPS models also apply to static models in most cases. Moreover, it is able to examine which metrics are needed to describe the real world driven by modeling.

To achieve this, we developed a meta-model of the ontology of metrics for modeling as the guidance of the research. Following the meta-model, we conducted a Systematic Literature Review (SLR) to aggregate, classify, and synthesize the metrics used in the SPS models. As a result, 145 studies that report SPS models are identified from the pool of SPS related papers until 2021. From the included studies, we extracted 2130 metrics. Although software metrics classification schemes have been proposed in the software measurement area~\cite{Akingbehin2006,Fenton2014Software}, they are not adaptive to SPS modeling since the focus of SPS and software measurement researches are different.

Under the guidance of the meta-model, we investigated metrics and their directly related elements from four research questions. We developed a new classification framework based on the extracted metrics and referred to the high level of categories (entities and attributes) suggested by the study~\cite{Fenton2014Software} (RQ1. metrics). Causal relationships between metrics in SPS models were discussed. We studied the considerations of metrics in terms of modeling purposes and paradigms of SPS models respectively (RQ2. causal relationships between metrics). Causal relationship diagrams that illustrate similarities and differences of modeling models at the cognitive level and models at tactical \& strategic levels are proposed and differences of metrics used in different paradigms are discussed (RQ3. selection of metrics). We provide a mapping on the relationships between data issues for measurement and solution strategies (RQ4. data for metrics). 

This study contributes to both the research community and practitioners in industry.
\begin{itemize}
\item It addresses the first challenge (identification of key metrics) in modeling; meanwhile, it provides researchers the necessary foundation for conducting research to solve the second challenge (data acquisition for measurement). 

\item This work serves as a knowledge base, enabling practitioners and researchers to gain a comprehensive understanding of the metrics and their relationships involved in the SPS modeling.

\item The classification framework and the considerations of metrics from modeling purposes, paradigms, and data issues provide a reference for practitioners to reuse existing knowledge; at the same time, it provides a clue to which metrics and causal relationships researchers need to focus on.

\item The second challenge is discussed based on evidence as a set of data issues, coping strategies, and available data sources for hard-to-get metrics are identified.

\end{itemize}

Note that we use a distinct citation format (author's surname \& year, e.g., \citesr{PfahlKR01a}) to distinguish the reviewed studies from other references. The complete list of references of the included studies is shown in the APPENDIX.

\section{Related Work}
This section introduces the process simulation and the software metrics that have been studied for decades in the SE community. 
\subsection{Software Process Simulation}
Kellner et al.~\cite{Kellner1999Software} offered an overview of the SPS area to answer three fundamental questions, i.e. \emph{why}, \emph{what}, and \emph{how}. They identified the reasons for conducting an SPS study, defined the scope of SPS models, and discussed the relationships among purposes, scope, and metrics. They also provided a framework to support decisions about simulation approaches, techniques, and required metrics.

Zhang et al.~\cite{ZhangKP08a,ZhangKP08b,ZhangKP10} conducted a series of SLRs on SPS modeling. They identify ten purposes for SPS research that are classified into three levels and two dimensions of the scope of the model~\cite{ZhangKP08a}. Several modeling paradigms and simulation tools are summarized in~\cite{ZhangKP08a,ZhangKP10}. They indicate five trends of process modeling based on the findings~\cite{ZhangKP08b}. The impact of SPS research on practice was also reported in another study~\cite{ZhangJHHZ11} with an impact roadmap that traces the successful SPS industrial application cases to their origins.

The impact of SPS has gradually expanded and it has become more and more mature in the last decade. The adoption of SPS in SE education is studied in~\cite{JiangZGSR15}, which confirms that education is an important application area of SPS with continuous research interests and shows that the SPS game appears more attractive to educators than the other forms of SPS. Integration strategies and recommendations for constructing hybrid SPS models are developed since software processes become more and more complicated~\cite{GaoZJ15}. The Verification and Validation (V\&V) take a critical role in securing the quality of SPS models, a mapping of quality aspects for V\&V and the possible V\&V methods is presented in~\cite{GongZYL17}.

However, there are still debates over the impact and usefulness of SPS research. Fran\c{c}a et al.~\cite{FrancaT13} present a quasi-systematic review of 108 studies to investigate the reliability of SPS studies in SE. As a result, they identified a few problems and indicated that SPS studies lack the necessary information for replication. 
Ali et al.~\cite{Ali2014A} aggregate all the points of view on the usefulness of SPS but find that no conclusion on these conflicting claims can be made based on the secondary studies. They conducted an SLR and evaluated 87 SPS studies. The results show that there is still a lack of conclusive evidence on it, as a few studies report the cost of developing an SPS model. Pfahl~\cite{Pfahl2014Process} also argues that it still lacks the evidence that SPS has become an accepted and regularly used tool for software project managers and its high cost is the main reason. 
On the other hand, the panel of domain experts on SPS collectively offered a different perspective that the impacts of SPS on practice cannot be ignored compared to many other SE technologies, although its high cost is still a major barrier against its wide application, and indicated that the consequences of waiving simulation should be considered whilst simulation is regarded as a cost saver rather than as a cost driver in other engineering disciplines~\cite{ZhangCrossroads14}.

\subsection{Software (Process) Metrics}
Software metrics play a crucial role in quantitative software engineering and have been researched for many years from different perspectives. We discuss the previous secondary studies of software metrics and present the overview in Table~\ref{tab:related work}.
\begin{table*}
\centering
\scriptsize
\caption{An overview of the comparison between this work and previous studies on software metrics} \label{tab:related work}
\begin{tabular}{p{26mm}p{5mm}p{55mm}p{77mm}}  
\toprule
\textbf{Studies} & \textbf{Year} & \textbf{Metrics' Classification} & \textbf{Research Focus} \\ 
\midrule
G{\'{o}}mez et al.~\cite{Gomez2006} & 2006 & Software Measurement Ontology~\cite{Garc2006Towards} & What, when, and how to measure  \\
Bellini et al.~\cite{Portobellini2008} & 2008 & Fenton's categorization~\cite{Fenton2014Software} & Measurement theory; Alternative methods to collect and analyze core measure; Metrics concepts, and how to identify metrics\\
Kitchenham et al.~\cite{Kitchenham2010WUS} & 2010 & OO metrics, web-metrics, and other code metrics & Identify trend in influential software metrics studies\\
Abilio et al.~\cite{Abilio2012A} & 2012 & Software attribute they measure in paper & Software maintainability metrics \\
Kupiainen et al.~\cite{Kupiainen2015} & 2015 & Fenton's categorization~\cite{Fenton2014Software} & Using metrics in Agile software \\
Nu\~{n}ez-Varela et al.~\cite{Nu2017Source}  & 2017 & Object oriented programming, aspect oriented programming, feature oriented programming, and procedural programming & The trend of source code metrics \\
Meidan et al.~\cite{Meidan18} & 2018 & Process, developer, project, and organization & Understand the measurement of the software development process \\ 

This work & 2022 & A detailed framework based on Fenton et al.’s categorization~\cite{Fenton2014Software} & Specific to metrics in software process modeling, including classification, causal relationships between metrics, selection of metrics in modeling, and data issues for measurement.\\

\bottomrule
\end{tabular}
\end{table*}

G{\'{o}}mez et al.~\cite{Gomez2006} performed a systematic literature review whose objective is to answer the questions of how, when and what to measure. 
They adopted the classification of concepts defined in the Software Measurement Ontology proposed by~\cite{Garc2006Towards} which aims to contribute to the harmonization of the different software measurement proposals and standards, providing a coherent set of common concepts used in software measurement. 

Bellini et al.~\cite{Portobellini2008} conducted an SLR to investigate five key conceptual and methodological issues, i.e. how to apply measurement theory to software, how to frame software metrics, how to develop metrics, how to collect core measures and how to analyze measures. They adopted Fenton's classification framework~\cite{Fenton2014Software}. They finally provide methods for collecting and analyzing the measurements and suggest that attention is increasingly being paid to multidimensional metrics. 

Kitchenham et al.~\cite{Kitchenham2010WUS} developed a preliminary mapping of software metrics studies that focus on identifying influential studies from 2000 to 2005 on software metrics. 
They conclude that empirical studies are of major importance to the software metrics research community. Although software metrics have been studied by a considerable number of researchers in various dimensions, the empirical methodology adopted needs to be refined.

Abilio et al.~\cite{Abilio2012A} presented an SLR to identify metrics associated with software maintainability and proposed for feature-oriented and aspect-oriented technologies from 11 primary papers. The metrics are classified according to the software attribute they measure, which ranges from architectural, parallel development, debugging, to quality attributes. 

Kupiainen et al.~\cite{Kupiainen2015} undertook an SLR on using metrics in industrial Lean and Agile software development. The authors classify the metric based on Fenton's classification framework~\cite{Fenton2014Software}. 
They identify the degree of influence and popularity of metrics in agile development and present a mapping between metrics and agile principles.

Nu\~{n}ez-Varela et al.~\cite{Nu2017Source} conducted a mapping study to investigate the trend of research on code metrics. They classified code metrics into four categories, including object-oriented programming, aspect-oriented programming, feature-oriented programming, and procedural programming.

Meidan et al.~\cite{Meidan18} conducted a mapping study to create a classification scheme of studies on the measurement of software development process in terms of source type, publication year, research type, contribution type, proposal type, validation type, entity/abstraction and study context. They identified 13 process attributes, 3 developer attributes, 4 project attributes, and 2 organization attributes.   

This study have different research scopes, and we focus more on process metrics, which can help SPS researchers understand using metrics in software process (simulation) modeling comprehensively.

\section{Ontology} \label{sec:ontology}
To systematically study metrics for SPS modeling, we proposed an ontology to illustrate the relationship between metrics and simulation modeling, as well as their related concepts. Our research revolves around this ontology.

\begin{figure}[ht]
\centering
 \includegraphics[width=0.5\textwidth]{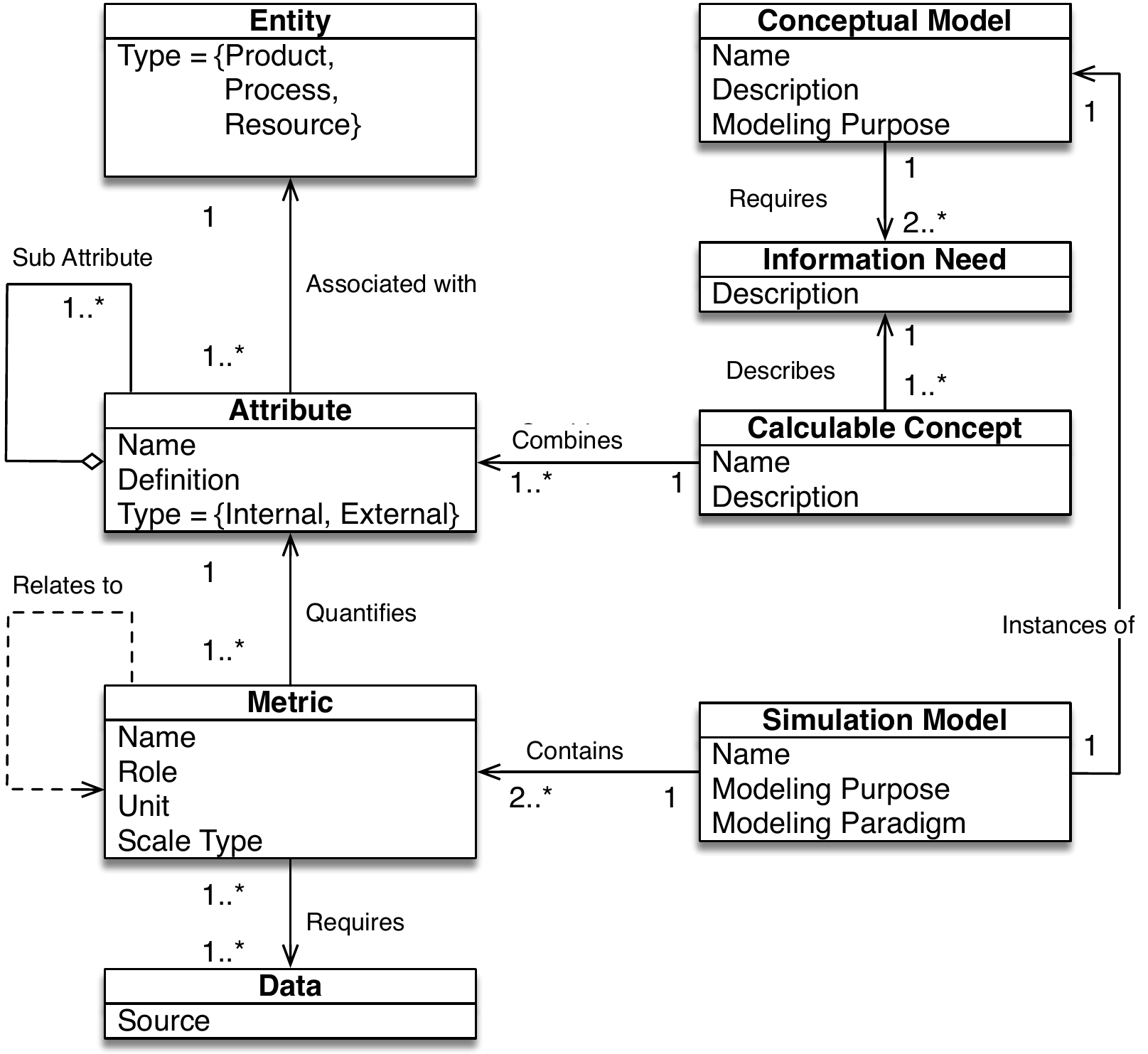}
\caption{Meta-model of the ontology for software metrics in software process simulation modeling} \label{fig:ontology}
\end{figure}

Olsina and Martín~\cite{olsina2003ontology} proposed an ontology for software metrics and indicators based on different software-related ISO standards and research articles. To adapt to SPS, we proposed an adjusted meta-model of the ontology referred to the metrics section of their ontology and introduced the concept of SPS modeling as presented in \figurename~\ref{fig:ontology}. The descriptions of the concepts and relationships presented in the meta-model are shown in Table~\ref{tab:concept description} and~\ref{tab:relationship description} respectively.

All models are simplified representations (abstractions) of the real world, as well as a conceptual model is the abstraction of a simulation model. All simulation modeling involves conceptual modeling~\cite{robinson2013conceptual}. Conceptual modeling is to specify a model that represents those parts of the problem domain that are included in the simulation model. According to the model development process suggested by Kitchenham et al.~\cite{KitchenhamPLJ05}, domain analysis and the definition of model requirements are needed before specifying a model. Domain analysis is to confirm the information needs in the real world. Not all information needs should be included in the model requirements since they should be valid, credible, feasible, and useful~\cite{robinson2008conceptual}, in other words, they are calculable concepts. A simulation model represents the conceptual model in a specific computer code, which means that simulation modeling is to quantify the input and output of the conceptual model. From the perspective of software metrics, it is to measure attributes of entities and quantify the causal relationships between attributes.

\begin{table*}
\centering
\scriptsize
\caption{Software metrics ontology: glossary of concepts} \label{tab:concept description}
\begin{tabular}{p{20mm}p{50mm}p{100mm}} 
\toprule
\textbf{Concept}& \textbf{Definition} & \textbf{Attribute: \textit{Description}} \\ 
\midrule
 \multirow{2}{*}{Entity} & \multirow{2}{*}{\shortstack[l]{Object that is to be characterised by measuring \\its attributes~\cite{standard2009adoption}.}} & Name: \textit{Name of an entity.} \\
   & & Description: \textit{An unambiguous description of the entity meaning.}\\
\hline   
 \multirow{3}{*}{Attribute} & \multirow{3}{*}{\shortstack[l]{A measurable physical or abstract property \\of an entity.}} & Name: \textit{Name of an attribute.} \\
   & & Definition: \textit{An unambiguous description of the attribute meaning~\cite{standard2009adoption}.}\\
   & & Type: \textit{Attributes can be internal or external.}\\
\hline   
 \multirow{5}{*}{Metric} & \multirow{5}{*}{\shortstack[l]{A metric is the number or symbol assigned to\\ an entity by this mapping in order to \\characterize an attribute.~\cite{Fenton2014Software} \\ Variable and parameter are synonyms of \\metric in an SPS model.}} & Name: \textit{Name of a metric.}\\
   & & Unit: \textit{Particular quantity defined and adopted by convention, with which other quantities of the same kind are compared in order to express their magnitude relative to that quantity.}\\
   & & Scale type: \textit{The type of scales depends on the nature of the relationship between values of the scale. These types of scales are commonly defined: nominal, ordinal (restricted or unrestricted), interval, ratio, and absolute.}\\
   \\
\hline   
 Data & The data that need to be collected in the real world for measurement. & Source: \textit{The source where the data can be obtained.}\\
\hline  
 \multirow{6}{*}{Conceptual Model} & \multirow{6}{*}{\shortstack[l]{A non-software specific description of the\\ computer simulation model (that will be, \\is or has been developed), describing the \\objectives, inputs, outputs, content, assu-\\mptions and simplifications of the model~\cite{robinson2008conceptual}.}} & Name: \textit{Name of a conceptual model.} \\
  & & Modeling Purpose: \textit{A specific purpose corresponding to the real world that why we develop a model.}\\
  & & \\
  & & \\ 
  & & \\
  & & \\
 \hline   
 Information Need & Insight necessary for the specific modeling purpose~\cite{standard2009adoption} & Description: \textit{An unambiguous textual statement describing the information needs.} \\
\hline  
 \multirow{2}{*}{Calculable Concept} & \multirow{2}{*}{\shortstack[l]{Abstract relationship between attributes of\\ entities and information needs~\cite{standard2009adoption}.}} & Name: \textit{Name of a calculable concept.} \\
  & & Description: {An unambiguous description of the calculable concept meaning.} \\
\hline
 \multirow{3}{*}{Simulation Model} & \multirow{3}{*}{\shortstack[l]{A simulation model is a computerized \\model that represents some dynamic\\ system or phenomenon~\cite{Kellner1999Software}.}} & Name: \textit{Name of a simulation model.}\\
  & & Modeling Purpose: \textit{A specific purpose corresponding to the real world that why we develop a model.}\\
  & & Modeling Paradigm: \textit{A distinct set of concepts or thought patterns for simulation modeling, e.g., System Dynamics.}\\

\bottomrule
\end{tabular}
\end{table*}

\begin{table}[ht]
\centering
\scriptsize
\caption{Software metrics ontology: relationship description} \label{tab:relationship description}
\begin{tabular}{p{18mm}p{60mm}} 
\toprule
\textbf{Relationship} & \textbf{Description} \\ 
\midrule
Sub-Attribute & An attribute may be composed of none or several sub-attributes, which are in turn attributes.\\
Associated with & One or more measurable attributes are associated with one or more entities.\\
Quantifies & One or more metrics can quantify an attribute.\\
Requires & It requires real-world data for measuring a metric. \\
Combines & A calculable concept combines (associates) one or more measurable attributes. \\
Consists of & A simulation model consists of a number of metrics.\\
Instances of & A simulation model is an instance of a conceptual model.\\ 
Relates to & One metric relates to another another metric. \\

\bottomrule
\end{tabular}
\end{table}

\section{Research Method}
This section describes the SLR process on metrics in SPS modeling that followed the SLR guidelines~\cite{keele2007guidelines}. Four researchers and their supervisor were involved in this study. 

Strictly speaking, this study is not an SLR research. We use the SLR research method to study metrics, which are usually not the research focus of the primary studies we included.


\subsection{Research Questions}
This study focuses on metrics and elements shown in the meta-model (\figurename~\ref{fig:ontology}) that are related to metrics. RQ1 studies metrics and their corresponding attributes from different perspectives. RQ2 studies the causal relationships between metrics presented in SPS models. RQ3 studies the purposes and paradigms of SPS models, which affects the selection of metrics. RQ4 studies issues and solution strategies of data for measurement. As an instantiated executable model of the conceptual model, the simulation model can cover the relevant information of the conceptual model. Hence, RQs are not specifically addressed to conceptual model, information need, and calculable concept that are involved in the meta-model. To achieve the research objective, four research questions are defined to drive this study following the meta-model (as follows).

  \textbf{RQ1: \emph{What metrics have been used and studied in the SPS studies?}}
  RQ1 aims to discuss what to measure in depth as well as build a classification framework of software metrics used and studied in the SPS models. As shown in the meta-model, metrics quantify attribute that is associated with entity. It will also be a multi-level classification of metrics that group them based on the attribute they measured and the entity corresponding to the attribute. In addition, the unit and scale type of each metric would be presented to show how to measure more specifically.

  \textbf{RQ2: \emph{What are the causal relationships between metrics used in different SPS studies?}}
  RQ2 helps to identify the common causal relationships among software metrics used in existing SPS models. The simulation model consists of causal relationships among metrics. The causal relationships between the metrics are the basis for understanding the structure of SPS models.
 
  \textbf{RQ3: \emph{What are the considerations for selecting metrics in terms of different modeling purposes?}}

  It is almost impossible to simulate all the factors and details in one model. Modelers select the most relevant metrics and ignore others as they concentrate on different purposes. Besides, modelers would adopt different modeling paradigms for different modeling granularities, which would affect the measurement of metrics. The goal of RQ3 is to investigate the considerations of metrics in terms of modeling purposes and paradigms, which are the two main properties of different simulation models as indicated in the meta-model.

  \textbf{RQ4: \emph{What are the issues and strategies for obtaining data for metrics.}}
  The lack of data remains a problem in SPS modeling, since Raffo and Kellner~\cite{RaffoK00} have analyzed different situations and solutions. RQ4 aims to revisit the status quo of data issues based on evidence, as well as investigate existing solution strategies and specific attributes they can measure.

\subsection{Selection Criteria}

The inclusion and exclusion criteria for the relevant studies are shown in Table~\ref{tab:selection criteria}. The included paper should be published in English and we retrieved the time span up to 2021. Our research included the studies that applied simulation modeling paradigms for software process research, software education, and software practice. Since the metric is our review focus, the selected studies should clearly claim the metrics used in their model. Regular papers can give us more comprehensive information to help us collect significant evidence, so we would not select the studies which have no more than 5 pages. Studies should be published as journal articles or conference papers rather than other forms that presented in C5 and C6. We extracted data from primary studies that could provide the first-hand metrics used in SPS research rather than from relevant secondary studies. Furthermore, evidence from primary studies could help us to summarize the relationships among different metrics in different SPS research. In addition, no secondary study was found that investigated the metrics used in SPS. To be specific, although Pfahl~\cite{pfahl2004pl} presents the example set of metrics used in the SD model which aims to support the analysis of the effectiveness of key SPI in the automotive industry (satisfy C1, C2, C3), the paper is excluded because it is a short paper (meet C4). In another example, Zhang et al.~\cite{Zhang2007A} mapped four typical simulation paradigms to the appropriate maturity levels of CMMI to adopt them; however, no specific model is presented in the paper (does not satisfy C2, C3). 

\begin{table}[ht]
\centering
\scriptsize
\caption{Selection criteria} \label{tab:selection criteria}
\begin{tabular}{p{85mm}} 
\toprule
\textbf{Inclusion criteria} \\
\midrule
C1. Published before 2021 and written in English. \\
C2. Primary studies on employing simulation modeling paradigms for software process research, education and practice.\\
C3. Primary studies that claimed the metrics used in the simulation model.\\
\hline
\textbf{Exclusion criteria} \\ \hline
C4. Short papers (no more than 5 pages).\\
C5. In the forms of editorial, abstract, keynote, poster, and book. \\
C6. Opinion pieces, comments, corrections, notes, slides alone or position papers. \\
C7. Secondary studies summarizing the outcomes of the existing research work, e.g. road-map, review, survey, etc. \\
\bottomrule
\end{tabular}
\end{table}

\subsection{Search \& Selection Process}

\begin{figure}[ht]
\centering
 \includegraphics[width=0.4\textwidth]{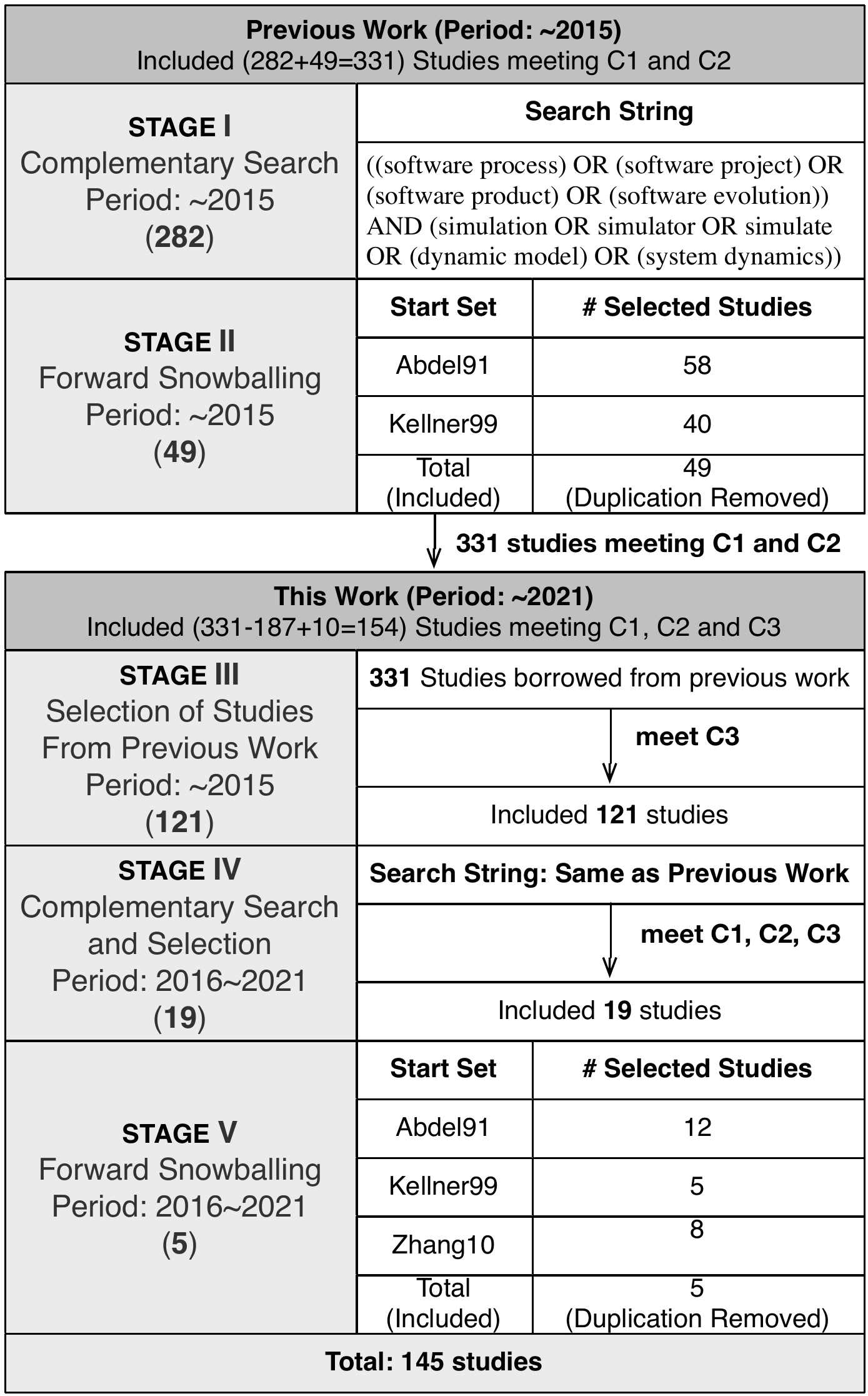}
\caption{Thorough literature search process} \label{fig:search process}
\end{figure}

\figurename~\ref{fig:search process} shows the search process of this study that consists of five stages.  Stages I and II were completed in our initial review~\cite{GaoZJ15}. In stage I, the manual search and the automated search were performed by two research students. The venues for manual search, which include five conferences and six journals, are listed in Gao et al.'s work~\cite{GaoZJ15}. The search string is shown in \figurename~\ref{fig:search process}. It was further coded into the equivalent forms to match the search syntax of different digital libraries. Four digital libraries (IEEE Xplore, ACM digital library, ScienceDirect, SpringerLink) were searched to retrieve as many SPS studies as possible. In stage II, the forward snowballing was applied as a supplementary~\cite{GaoZJ15}. As a result, a total of 331 candidate studies were identified by scanning the title, keywords, and abstract.

The only difference in the selection criteria between this study and the study by Gao et al.~\cite{GaoZJ15} is C3, which identifies the studies reporting modeling metrics (instead of the hybrid simulation modeling in the previous study~\cite{GaoZJ15}) from all relevant SPS studies. Consequently, in stage III, the 331 candidate studies were checked against the selection criteria by further reading the introduction, conclusion, and even full text iteratively until the final consensus was reached. Candidate papers were assigned to four research students and each paper was assigned to at least two students to enable parallel review. All inconsistencies and disagreements were thoroughly discussed and resolved during weekly meetings with supervisors. In this study, only the latest versions of primary studies were selected for review if different versions were published based on the same model.

In stage IV, we replicated the automated search using the same search string and extended the search scope to 2021. We identified 19 papers that met the selection criteria published between 2016 and 2021.

In stage V, we conducted forward snowballing using Google Scholar. The seminal set for the forward snowballing include ``Abdel91''~\cite{Abdel1991Software}, ``Kellner99''~\cite{Kellner1999Software}, and ``Zhang10''~\cite{ZhangKP10}, which were cited by most relevant studies from the early stages of the review. We identified 5 studies after deduplication. 

\subsection{Data Extraction} \label{sec:extraction}

\begin{table}[ht]
\centering
\scriptsize
\caption{Data extraction scheme} \label{tab:data extraction}
\begin{tabular}{p{22mm}p{45mm}p{7mm}}  
\toprule
\textbf{Item} & \textbf{Description} & \textbf{RQ(s)} \\
\midrule
Title & The title of the study.& Info\\
Year & The published year of the study. & Info\\
Paradigms& System Dynamics; Discrete-Event Simulation; Agent-Based Simulation; Hybrid simulation, etc. & Info\\
Software metrics & The metrics involved in the SPS studies, including their names, units, and original descriptions.& RQ1-4 \\
Causal relationships between metrics & the predecessor metric that output to or affects the current metric, and the successor metric that is inputted from or affected by the current metric. &RQ2\\
Modeling purposes & The reasons form SPS models.&RQ3\\
Data issues & The issues of selecting metrics due to data availability. & RQ4\\
Solutions of data issues & The strategies that solve the data issues. & RQ4\\
Data source & The source of data related to solutions for measuring attributes, one study may use multiple data sources. & RQ4\\

\bottomrule

\end{tabular}
\end{table}

To answer the research questions, we defined a data extraction scheme to collect important information from the reviewed studies. Each research question is answered by at least one extraction item. As shown in Table~\ref{tab:data extraction}, the data extraction scheme includes the citation information (e.g., title, year and modeling paradigms) and the information specifics to the research questions. Software metrics (names, units and original descriptions) are extracted for answering RQ1,2,3,4. Causal relationships between (two) metrics and modeling purposes are identified for RQ2 and RQ3, respectively. Data sources and issues and solutions are identified for RQ4.

We started with a pilot extraction, in which 35 randomly picked papers was allocated to all researchers. We noticed that some of the papers clearly introduced metrics used for their SPS modeling or presented the SPS model. For these cases, we can easily identify the software metrics and their causal relationships. Some papers did not present software metrics explicitly; the metrics and causal relationships might be identified from its context by iteratively reading the full text. As a result, we extracted 2130 metrics and identified 183 types of causal relationships from the 145 identified papers. These metrics, which are also called variables, comprise the SPS models.

\subsection{Data Synthesis \& Classification} \label{sec:synthesis}

\begin{table*}
\centering
\scriptsize
\caption{Example of coding} \label{tab:coding}
\begin{tabular}{p{70mm}p{30mm}p{30mm}p{20mm}p{10mm}}  
\toprule
\textbf{Metric} & \textbf{Code-I1} & \textbf{Code-I2} & \textbf{Code-I3} & \textbf{Code-I4}\\ 
\midrule
\emph{E1:} Residual defect density (i.e. actual reported defects that were not corrected after 1094 days) & Defect, Density, Residual & Defect, Density, Remaining & Defect, Density & Defect \\
\emph{E2:} Number of tasks completed & Task, Number, Completed & Task, Number, Completed & Task, Size & Task\\
\emph{E3:} Delay from the completion of this until next release is delivered to users & Delay, Release & Delay, Release & Delay, Release & Time \\

\bottomrule

\end{tabular}
\end{table*}

We applied the thematic synthesis method~\cite{Cruzes2011Recommended} to construct our findings in a systematic manner. As the metric descriptions vary significantly between process modelers, to answer RQ1, we synthesized the list of the extracted software metrics using the coding technique~\cite{Corbin1998Basics}. It is an iterative process to develop a consistent set of codes from the diverse descriptions in the reviewed studies. Table~\ref{tab:coding} shows three examples of the evolution of codes. In the first iteration, two researchers coded all the metrics at a detailed level based on their descriptions independently, then we discussed the differences between the two sets of codes, and finally reached agreement. In the second iteration, we identified similar codes in the Code-I1 which developed in the first iteration and replaced them into the same code, for example, we coded `issue', `error', `fault', `flaw', `bug' as `defect'; coded `fix', `fixed', `correct', `correction' as `fixing'. For the Example 1 (E1), the `residual' is replaced by `remaining' from Code-I1 to Code-I2. Some codes have no synonyms that could be replaced; then Code-I2 would be the same as Code-I1, e.g., E2. We gradually made the code more abstract in the third and fourth iterations through discussion. Some codes did not change from I2 to I3 to replace few similar metrics, e.g., E3. We developed Code-I4 based on Code-I3 and partially referred to the Fenton et al's software metric taxonomy~\cite{Fenton2014Software} which has been widely accepted in the community. As a result, all metrics can be classified into 29 different categories according to Code-I4, which can be grouped by product internal/external, process internal/external and resource internal/external as suggested by the study~\cite{Fenton2014Software}. The categories were classified into sub-categories according to Code-I3.

In the process of coding-based classification of metrics, metrics that measure time, effort, and cost, etc. can be easily identified and classified. But we have also encountered some thorny problems, which are not covered in Fenton et al's work~\cite{Fenton2014Software}. There are a variety of factors and multipliers that affect other diverse variables in the model. We grouped the factors into two major categories, i.e. Process factor and Manpower factor, based on the variables that are affected by them. For example, various policies were modeled in existing SPS models; we classified \textit{pair programming policy} which is a boolean variable that determines whether to apply pair programming in the Process factor; and classified \textit{staffing policy} which determines the number of person months assigned to the Manpower factor. In Fenton et al.'s framework~\cite{Fenton2014Software}, \textit{defect} is the only category related to defects, and they suggested that \textit{defect} should be the Process attribute. In our opinion, although defects are generated and fixed in the process, defects that exist at a certain point in time exist objectively as part of the product. Therefore, we distinguish \textit{defect activity} from \textit{defect} and classify \textit{defect} into Product category as it is more reasonable. We classified metrics such as \textit{defect fixing rate} into \textit{defect activity} which belongs to Process and classified metrics such as \textit{\# fixed defects} into \textit{defect}. Similarly, we perform the classification based on the results of coding (i.e. the abstract meaning of metrics) on the one hand and on the other hand, based on the context of the metrics. We refer to existing classifications, but we are not bound by them.

The content (frequency) analysis was performed throughout the study to analyze the nominal data across groups of variables. To identify the most common causal relationships between metrics, the percentage of the occurrence of causal relationships was counted when answering RQ2.

For RQ3, we counted the frequencies of different categories of metrics used for two hierarchical levels of modeling purposes as well as for different paradigms to locate the evidence that leads to the difference so that we could conduct an in-depth analysis.

\section{Results} \label{sec:results}
This section presents the distribution of included studies from three aspects, i.e. publication years, paradigms used in studies and modeling purposes.

\subsection{Years}
The review identifies 145 papers after the five stages (as presented in \figurename~\ref{fig:search process}). The selected papers for review were published from 1997 to 2021. The complete list of references of the included studies is shown in the APPENDIX.

\begin{figure}[ht]
\centering
 \includegraphics[width=0.48\textwidth]{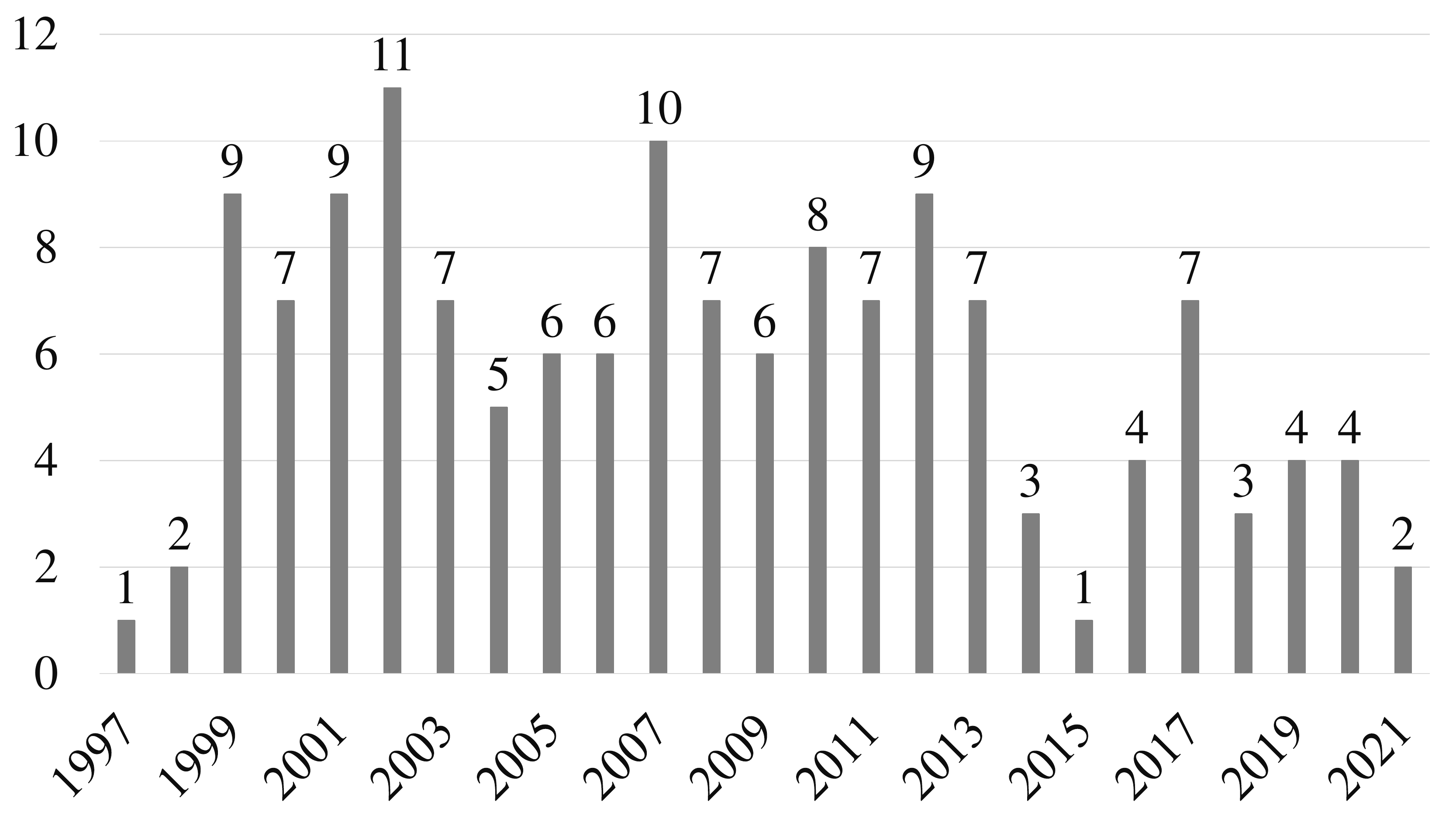}
\caption{Study distribution per year} \label{fig:year}
\end{figure}

\subsection{Paradigms}
\begin{figure}[ht]
\centering
 \includegraphics[width=0.38\textwidth]{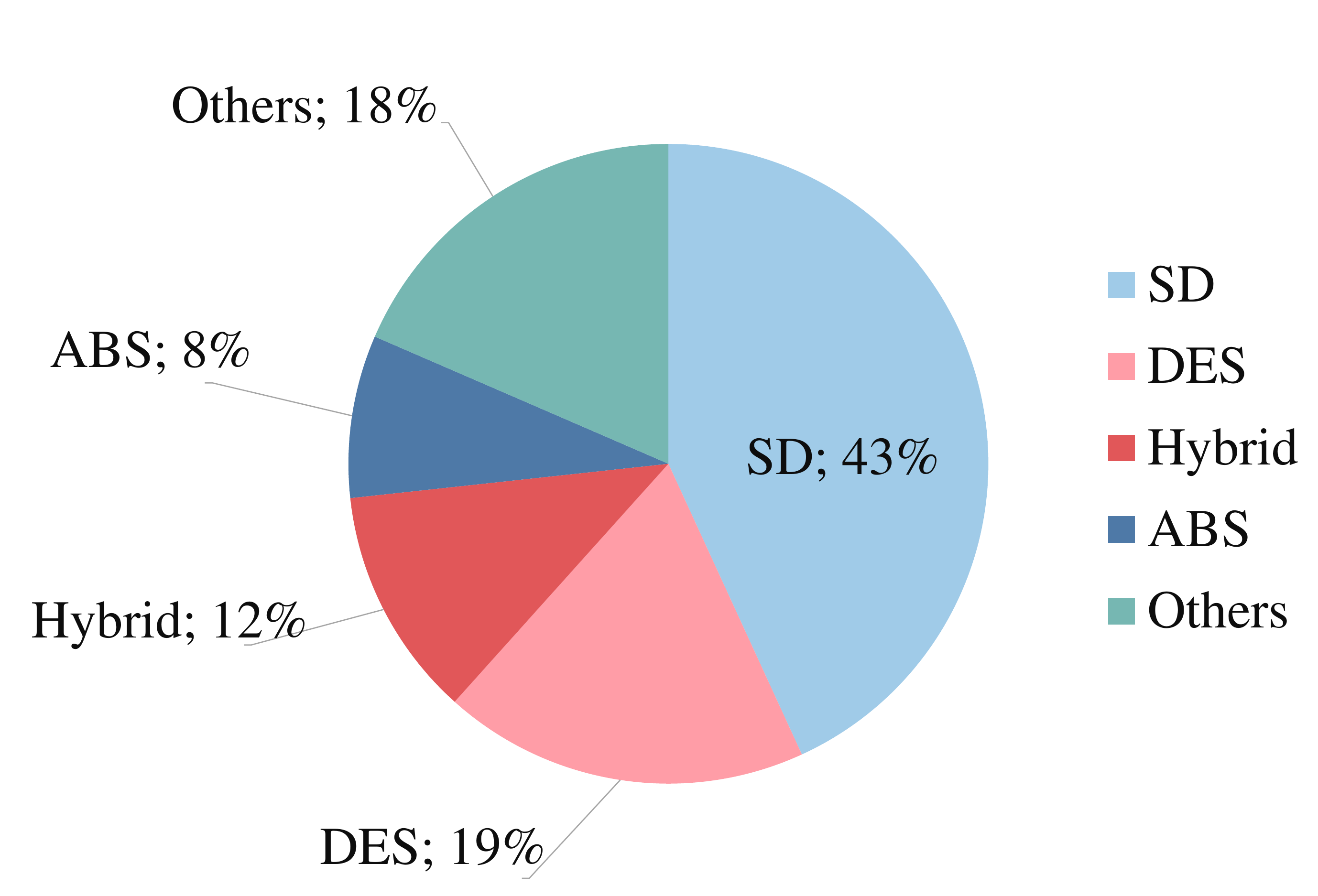}
\caption{Study distribution per modeling paradigm} \label{fig:modeltype}
\end{figure}

We classified the modeling paradigms applied in the SPS studies as shown in \figurename~\ref{fig:modeltype}. System Dynamics (SD, 43\%) is the most popular modeling paradigm in SPS. The rest includes Discrete-Event Simulation (DES, 19\%), Hybrid Simulation (Hybrid, 12\%) and Agent-Based Simulation (ABS, 8\%). A hybrid model may adopt two or more modeling paradigms together. Other studies use a variety of modeling paradigms that simulate software processes at different abstraction levels distinct from the above, e.g., Qualitative Simulation (QSIM), Parametric Estimating (PE), etc.

\subsection{Modeling Purposes} \label{sec:purposes}

\begin{figure}[h]
\centering
 \includegraphics[width=0.49\textwidth]{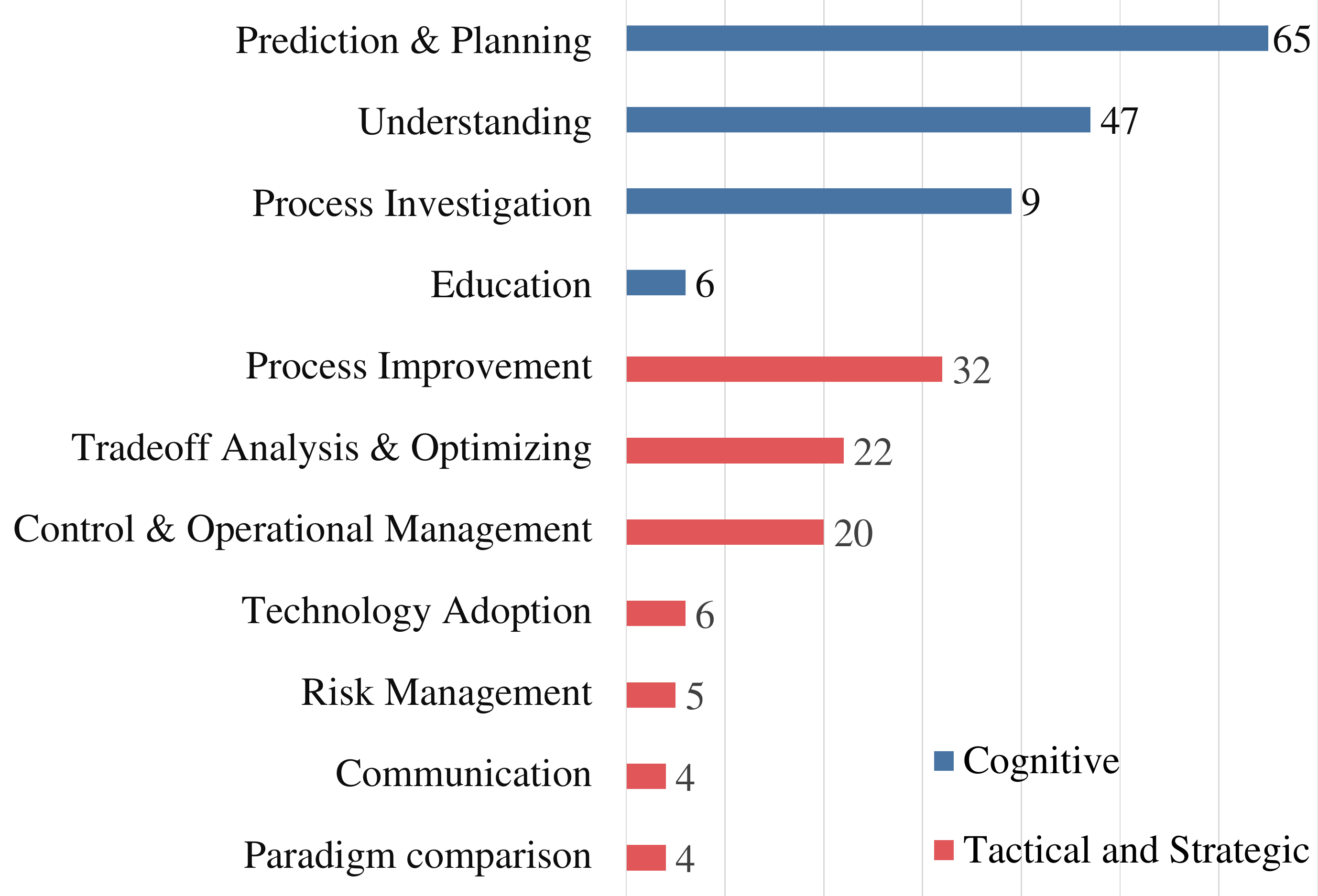}
\caption{Study distribution per modeling purpose (One study may have multiple modeling purposes)} \label{fig:purpose}
\end{figure}

Kellner et al.~\cite{Kellner1999Software} identified the reasons for SPS models and clustered them into six specific modeling purposes to perform the simulation of software processes. Zhang et al.~\cite{ZhangKP10} extended it to ten purposes based on the systematic review of published SPS studies. In their study, these purposes were grouped into three levels, i.e. cognitive, tactical \& strategic. The cognitive level includes understanding, communication, process investigation, and education. The rest of the purposes are at both tactical \& strategic levels. 

We grouped the modeling purposes of the studies into eleven purposes (we identified a new modeling purpose). Different from the other ten purposes, the new purpose, paradigm comparison, is related to the modeling paradigm, but not for studying the software process. The purpose is to compare the strengths and weaknesses of different paradigms. In the study whose goal is to compare paradigms, the paradigm comparison is the sole purpose of building the models. For example, a qualitative and a quantitative model of the typical software evolution process were built for comparing SD with qualitative modeling diagram and there is no redundant discussion about the value of the models themselves~\citesr{Zhang2009Qualitative}. The models built for paradigm comparison are simple but complete enough which are based on the degree to which the simulated behaviors interpret the process. From this point of view, these models are at the cognitive level. 

As shown in \figurename~\ref{fig:purpose}, we grouped 42 studies into the cognitive level and the remaining 103 into the tactical \& strategic level. One study may have multiple modeling purposes, therefore, one study may have both cognitive and tactical \& strategic modeling purposes. For such research, we consider its modeling granularity to be at the tactical \& strategic level. Understanding and process investigation are the most common modeling purposes at the cognitive level. Prediction \& planning and process improvement are the most common modeling purposes at tactical \& strategic level.

\section{Findings}
Kaner et al.~\cite{Kaner2004Software} collected the definitions of measurement, a concise definition they recommend is provided by Fenton and Pfleeger as below~\cite{Fenton2014Software}. 

\textit{Formally, we define measurement as a mapping from the empirical world to the formal, relational world. Consequently, a metric is the number or symbol assigned to an entity by this mapping to characterize an attribute.}

In SPS modeling, a metric is also called variable sometimes, in our opinion, metric is a more appropriate name since it implies the mapping from real software process to the SPS model. However, the boundary between attribute and metric is not strict in SPS studies, e.g., \textit{size of code} and \textit{number of defects} are the common names of two variables, however, the former is an attribute and the latter is a metric according to the definitions in the software measurement area. The metric of \textit{size of code} should be \textit{lines of code}. This kind of ambiguity will cause trouble when we classify them. Hence, we treat all variables as metric and keep their original representation as much as possible.

Although there exists a body of knowledge on software metrics~\cite{Fenton2014Software}, no systematic and comprehensive research on metrics for process modeling has been reported in the SE community. The selection of metrics in SPS models turns out to be more challenging than static models because of the dynamic nature and the extra requirements for executability. Therefore, this study concentrates and reports on the metrics used and studied in SPS modeling only.

\subsection{Metrics and Classifications (RQ1)} \label{sec:rq1}

We extracted a total of 2130 metrics used (or studied) in SPS models from the 145 reviewed studies. There are many identical or similar metrics in the data set; hence, we classified them for analysis. Based on the meta-model, our classification framework would contain four levels of categories, from abstract to concrete, these are entities, attributes, sub-attributes, and metrics.

\subsubsection{Categories of Entities}

We refer to categories and definitions in Fenton et al.'s classification~\cite{Fenton2014Software}. We classify software metrics into three categories as \emph{products}, \emph{processes}, and \emph{resources} (aligned with a Goal-Based framework for software measurement), where \emph{processes} are activities that evolve over time and have to be completed in sequence during development, \emph{products} are generated from process activities including artifacts, deliverables, and documents, and \emph{resources} are the entities needed in performing activities. 

In each category, we further distinguished two types of metrics, i.e. \emph{internal} and \emph{external} metrics. An \emph{internal} metric can be measured only by the intrinsic properties of \emph{product}, \emph{process}, or \emph{resource} on its own without considering their observable behaviors. On the contrary, an \emph{external} metric is measured purely by taking impacts that it may make on the \emph{product}, \emph{process}, or \emph{resource} into account. 

\subsubsection{Categories of Attributes}
There is a large amount of studies in software metrics/measurement, however, the roles of metrics in SPS studies are different from the roles in them, which would easily make modelers get confused. For example, the software measurement related studies are full of research on coupling and cohesion whilst these are rarely used in existing SPS models. It implicates that existing classification frameworks, which were built based on software measurement research and knowledge, are not suitable for SPS modeling. Furthermore, one of the most recognized frameworks, Fenton et al.'s classification framework, does not fully cover the metrics used in modeling. In the study \cite{Fenton2014Software}, only examples of attributes are presented for each entity, which is far from enough to guide modeling. 

In this study, we built a new classification framework of metrics used in SPS models following the thematic synthesis method described in Section~\ref{sec:synthesis}. The framework is divided into four figures, i.e. \figurename~\ref{fig:metrics-product}-\ref{fig:metrics-resource}, due to the space constraints of one page. In these figures, the first column shows whether the measured attributes belong to an internal or external entity. The numbers in brackets indicate the number of metrics that measure these attributes. The second and third columns show the attributes and their sub-categories. The last column lists typical metrics with their units. Each metric consists of two parts; the upper rectangle shows the name of the metric and the lower rectangle shows its unit. For example, the first metric in \figurename~\ref{fig:metrics-product} is \textit{size of code} and its unit can be LOC, DSI or Function Points (FPs). The height of the rectangle of a metric is within the scope of the sub-category it belongs to, e.g, \textit{size of code} measures \emph{artefact size}, specifically, code size. It should be noted that \textit{size of code} should be a name of attribute from the semantic point of view, however, these are the minimum units in SPS models. We name these metrics following the coding technique and comply with the original expression of them in reviewed studies. To achieve the goal of modeling, various means (indicate by units) of measuring a specific metric were used in different studies. It is largely determined by the modeling object and the method of measurement that may lead to different units of the metric. Separating all similar metrics (e.g., both LOC and DSI are the metrics of code size) will make the results and discussion too trivial. Hence, we leave out some of the non-essential results.

The unit also implies the scale type of a metric, especially for numerical data. As shown in Table~\ref{tab:concept description}. Scale types are nominal, ordinal (restricted or unrestricted), interval, ratio, and absolute according to ISO/IEC 15939. The absolute and ratio are numerical data, which can be indicated by units. For instance, \textit{\# defects} is absolute and \textit{defects \%} is ratio. Besides, the type name of boolean, interval, and ordinal types are presented instead of presenting the unit since these metrics are dimensionless (i.e. there is no specific unit). For those metrics with a clear ordinal set or range, the ordinal set or range is also presented, e.g., \textit{expertise of developer} commonly have three levels {0,1,2} and \textit{knowledge level} have a finer level of granularity ranged from 0 to 100.


The classification framework presented in the figures is elaborated from \emph{product}, \emph{process}, and \emph{resource} categories as follows.

\subsubsection{Product Metrics}
\begin{figure}[ht]
\centering
 \includegraphics[width=.5\textwidth]{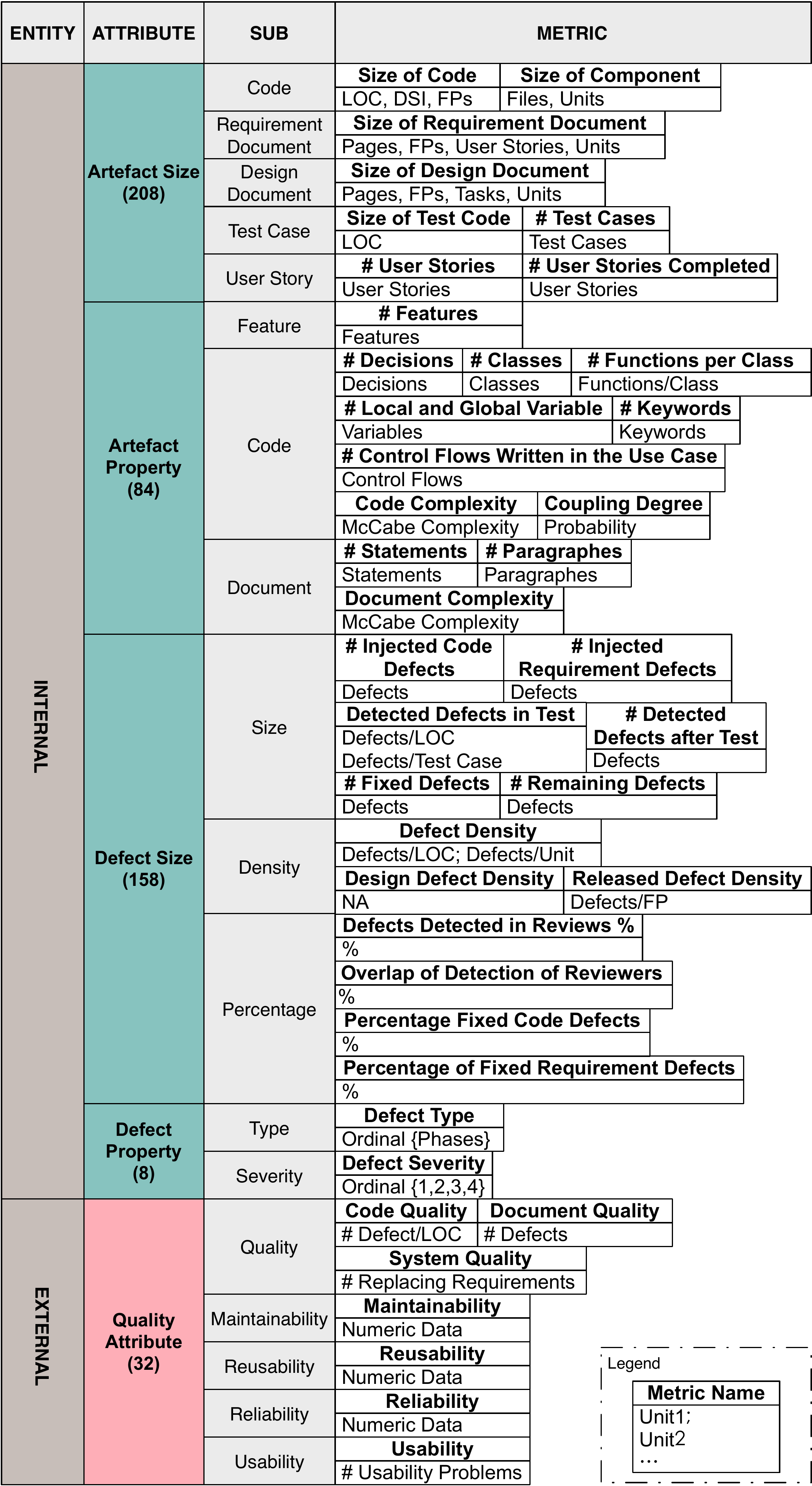}
\caption{Classification of Product Metrics} \label{fig:metrics-product}
\end{figure}

As shown in \figurename~\ref{fig:metrics-product}, the internal metrics are further divided into three sub-categories, i.e. \emph{artefact size}, \emph{artefact property} and \emph{defect}. 

\textbf{\emph{Artefact Size}} measures the amount of work product to be produced from the process. Generally, size is used to measure the scale of a project and further compute indirect attributes. Metrics for various types of artefacts were used in different studies, the collected metrics can be classified into five sub-categories as shown in \figurename~\ref{fig:metrics-product}. \textit{Size of code} that measures program size is most used metric in all the size metrics and is associated with the effort of development and maintenance and the faults generated~\cite{Fenton2014Software}. Meanwhile, \textit{size of requirement document}, \textit{size of design document}, and \textit{number of (\#) test cases} are widely used in SPS models. In detail, test cases are generated by either manual or automated approaches, while the latter is in consideration of the tools supporting model-based testing and made by software specifications written in more formal notation or structure~\citesr{Aranha2008Using}. In some studies such as~\citesr{WernickH02,Zhang2012Toward}, the artefact size is abstracted into a number of arbitrary-sized ``units'' in order to represent some suitable measure to the size of the system in reality. The \textit{\# user stories} and \textit{\# features} are used in different project contexts.

\textbf{\emph{Artefact Property}} is used to measure products at a more detailed level; e.g., code can be further measured by \textit{\# decision statements in the code}, \textit{\# local and global variables}, document can be measured by \textit{words per page}, etc. To be specific, \textit{\# control flows identified in the skeleton} is used to estimate the effort required to perform the requirements analysis activity in a DES model~\citesr{Aranha2008Using}. \textit{Coupling degree} and \textit{code complexity} also belong to this category. The former is used only in one study and is measured by the probability that one component will be affected by another~\citesr{Padberg2011Model}. The latter can be measured in different ways. Smith et al.~\citesr{Smith2006Agent} used McCabe complexity to measure code complexity. Raffo et al.s~\citesr{RaffoHV00} suggest that \textit{number of decision statements} in the code can be the metric of code complexity, \textit{number of local and global variables} and \textit{level of control-flow nesting} are the alternative metric. \textit{Flesch-Kincaid Grade Level Score} and \textit{Gunning Fog Index} are suggested as metrics of document complexity.

\textbf{\emph{Defect Size}} measures the amount of defects that are injected, detected, and may remain throughout the development process. According to the life cycle of defects, they roughly include injected, detected, fixed, and remaining defects. From the perspective of measurement, the metrics are classified into size, density, percentage, and property sub-categories. The \textit{\# defects} is the most frequently used type. The \textit{defect density} and \textit{percentage of fixed defects} evaluate the quality of the product from two different angles, the former evaluates from the product itself and the latter evaluates from the progress of fixing defects.

\textbf{\emph{Defect Property}} is used to measure defects at a more detailed level. Properties such as \textit{type} and \textit{severity} are introduced in several studies~\citesr{IOANA2003SUPPORTING,ZhangJZ08,Cangussu2008Using,lunesu2021assessing} to make the models behave closer to reality. The \textit{type} can be the ordinal data that indicates the phase of the defect injected~\citesr{RusBH02}. The \textit{severity} can be ordinal data with three to four levels, e.g., four levels of defect severity are modeled (i.e. ``easy'', ``medium'', ``hard'', and ``very hard'') by Zhang et al.~\cite{ZhangJZ08}.


The \textbf{\emph{external metrics}} of the product were not commonly used in SPS. They measure the \textbf{\emph{quality attributes}} of product such as \textit{reusability}, \textit{maintainability}, \textit{quality}, etc. While \textit{quality} is the commonly used metrics in this sector, the means to measure \textit{quality} of code and document vary from model to model. Pfahl and Lebsanft adopted an SD model to analyze the impact of software requirement volatility, the \textit{quality of the system} is measured by the number of replacing requirements~\citesr{PfahlL00}. From business value concerns, Madachy~\citesr{Madachy2005Integrated} measured the \textit{quality} by the number of defects. Only one study used the \textit{maintainability} metric and measured it using the equation $Maintainability = 13.12 * Complexity + 0.17 * Effort + 3.87 * Size$~\citesr{Ara2012Towards}. The \textit{reusability} and \textit{reliability} are commonly based on COCOMO~\uppercase\expandafter{\romannumeral2}, they are used as cost drivers with a nominal value of 1~\citesr{uzzafer2013simulation,Silva2013Decision}.

\subsubsection{Process Metrics}
\begin{figure*}
\centering
 \includegraphics[width=.93\textwidth]{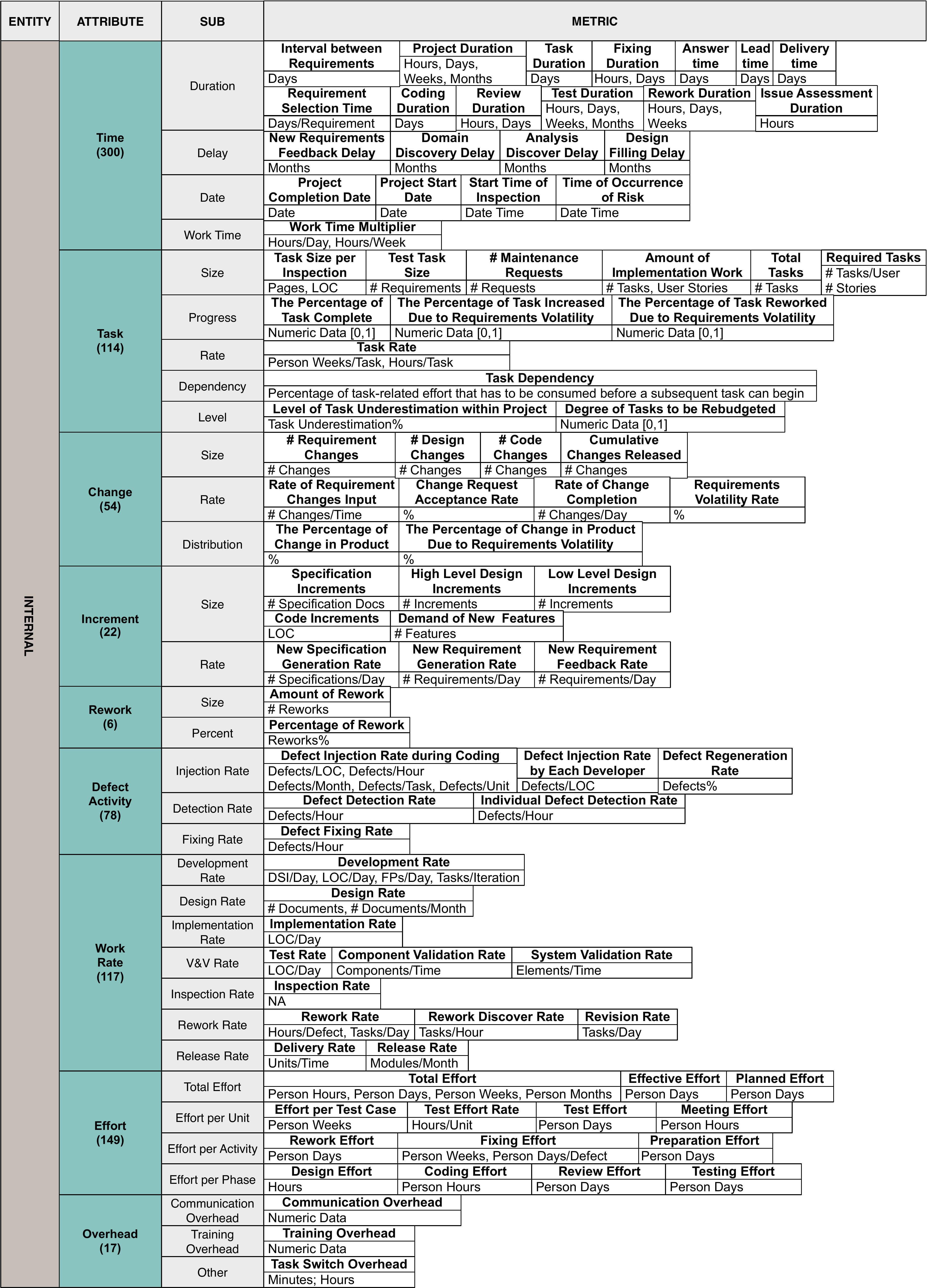}
\caption{Classification of Process Metrics (Part1)} \label{fig:metrics-process-a}
\end{figure*}

\begin{figure*}
\centering
 \includegraphics[width=.93\textwidth]{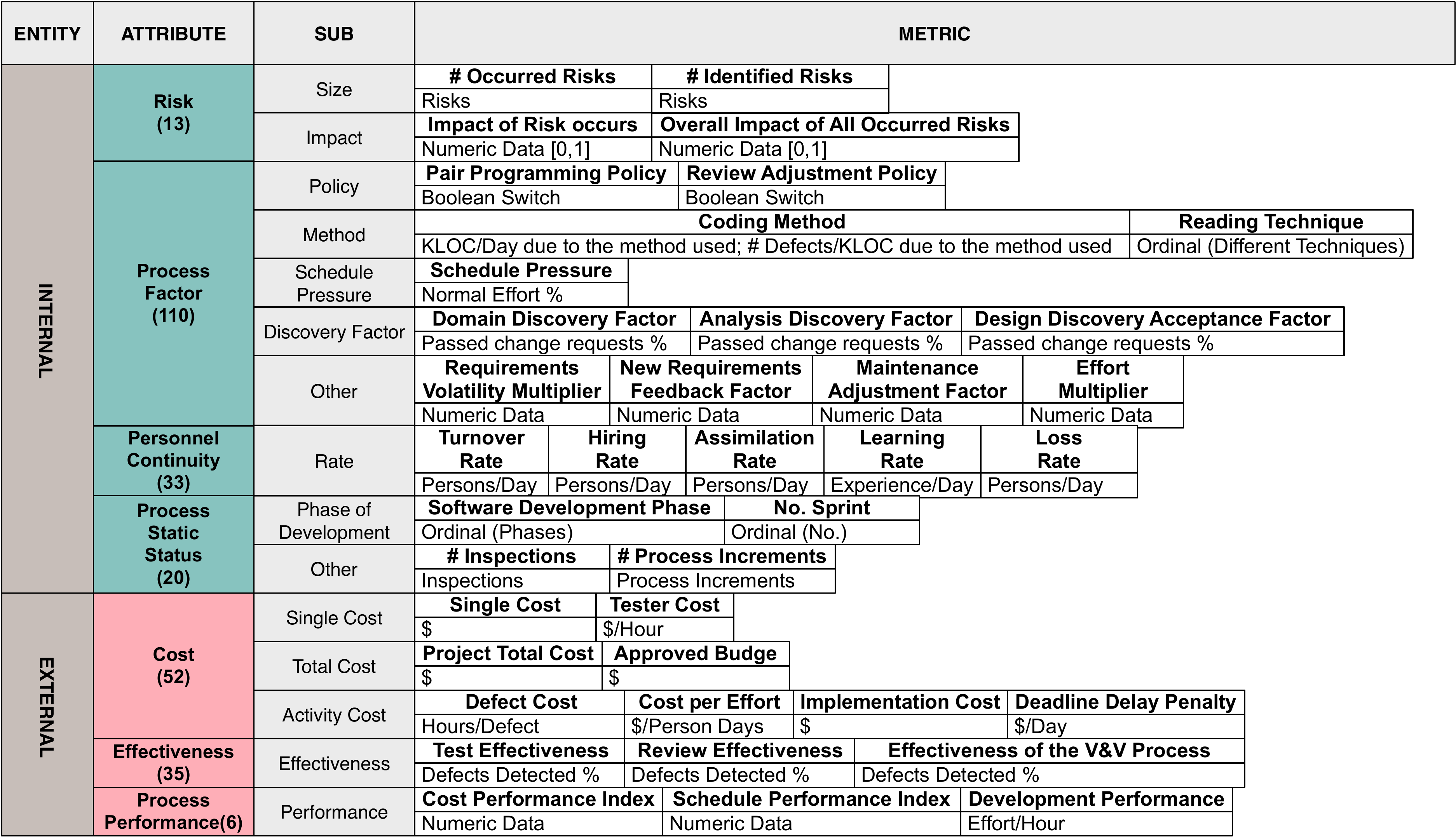}
\caption{Classification of Process Metrics (Part2)} \label{fig:metrics-process-b}
\end{figure*}

With the focus on process modeling, \figurename~\ref{fig:metrics-process-a},~\ref{fig:metrics-process-b} indicate that it is a large set of process-related metrics that can be further classified into a number of categories such as \textit{time, task, effort}, etc.

As a distinction between dynamic process models and static process models, there are four main types of \textbf{\emph{time metrics}}. The first is \emph{duration} which denotes the time spent on a single phase or the entire process. The model simulates a real process over a period of time, no matter whether \emph{duration} is an explicit metric indicated in an SPS model. \emph{Calendar date} could be an alternative metric to \emph{duration} when the model has many overlapped duration metrics. In a stochastic simulation model for risk management, \textit{the start dates and end dates of every risk} is measured~\citesr{Zhou2012A}. The third type is \emph{work time} which is often used as a multiplier in the model to simulate \textit{\# hours work per day}~\citesr{WakelandSR05}. \emph{Delay} is the last type of time metrics such as \textit{hiring delay}, \textit{delay from completion to release}, \textit{new requirements feedback delay}, etc.

As an alternative metric of artefact size in SPS, \textbf{\emph{task size}} runs through phases of the development process as an indicator of the job size of the corresponding process (phase). 

\textbf{\emph{Change}} occurs during requirement, design, implementation, or throughout the entire process. It is one of the major reasons for \emph{rework}. From the reviewed studies, this kind of metric is not widely used in SPS models.

\textbf{\emph{Increment}} represents the process that the new requirements or increments are adding into the development process. Both \textit{\# new requirements} added and the \textit{new requirement generation rate} are metrics that measure this process.

\textbf{\emph{Rework}} is the feedback of \emph{change} and \emph{defect activity}. The \textit{amount and percentage of task that required rework or were reworked} are \emph{rework} metrics. We classified the \textit{rate of rework} into the \textit{work rate} category to emphasize the activity and the `speed' of the process.  

We introduced the \textbf{\emph{defect}} in product entity. There are also some defect related metrics should be process metrics, we classified them into \emph{defect activity}. Metrics in \emph{defect} measures the static attributes of product defect, such as defect density. Metrics in \emph{defect activity} measures the attributes of dynamic activities, e.g., \textit{defect injection rate during coding}. The rate can be measured by the number of defects injected per time unit, also can be measured by the number of defects per artefact size (LOC).  

\textbf{\emph{Work rate}} represents the `speed' of activities are executed, which covers the entire life-cycle, including the processes of design, implementation, review, testing, releases, etc. \textit{Development rate} is the most common one, which measures the primary activity simulated in a model. Only using the \textit{development rate} would regard the process as a whole and will not distinguish the speed of different activities in the process. On the contrary, the speed of different activities can be measured respectively by \textit{design rate}, \textit{implementation rate}, etc.

\textbf{\emph{Effort}} is used (or studied) in numerous models as it is a special interest of process modeling. Its metrics appear as for example \textit{design effort}, \textit{rework effort}, or even \textit{total effort}, in corresponding phases except requirements. The effort can be typically measured with \emph{person-hours}, \emph{person-days} and \emph{person-months} which depends on the resolution of the model. In software process research, \emph{Effort} is regarded as the major contributor to \textit{cost} or its alternative~\citesr{RusNM14}.

\textbf{\emph{Risk}} rarely appears in SPS models. In the model that aims to investigate the impact of risk, the number, impact and occurrence time of risks are used in the model~\citesr{Zhou2012A}.

\textbf{\emph{Overhead}} results from the situation where extra work beyond the tasks cannot be avoided. \textit{Communication overhead} is the most common type of overhead. We classify them into process category because it is generated from the communication activity (or other activities) in teamwork. Other kinds of overhead such as \textit{task switch overhead} and \textit{training overhead} are also considered in several studies~\citesr{Baum2017Comparing,GarousiP16}.

\textbf{\emph{Process factors}} that generally denote the managerial and organizational factors make impacts through the whole process. In SPS models, \textit{policies} are the most metrics in this subcategory but vary significantly among different models. \textit{Schedule pressure} is a common factor of \textit{productivity} or \textit{work rate}. Various factors were used in studies, which are difficult to exhaust here. Various of \textit{discovery factors} which determines the proportion of elements which might possibly progress to the next stage can be calibrated using historical data~\citesr{WernickH02}. The determination of what factors should be considered depends on specific project characteristics and organizational characteristics.

\textbf{\emph{Personnel continuity}} relates to the training and turnover process and metrics measure the `speed' of these processes. For example, to model the human resource evolution process, the \textit{hiring rate}, \textit{dismissal rate}, \textit{turnover rate} can be considered~\citesr{RuizRT01}. 

\textbf{\emph{Process static status}} includes the metrics that are predefined and do not change during the modeled process, such as \textit{stage of the development} life-cycle, \textit{\# process increments}, etc. The \textit{stage of the development} can be combined with product features such as \textit{system type} to measure the \textit{capacity}~\citesr{HurtadoROT15}. 

\textbf{\emph{Process external metrics}} are occasionally observed in the reviewed models. They include \textit{effectiveness}, \textit{cost}, and \textit{process performance}, in which \textit{effectiveness} and \textit{cost} were used a lot. For software process, there are many concerns that can be defined as the indicators of \emph{effectiveness}, such as \textit{test requirement defect detection effectiveness}~\citesr{Cangussu2008Using}. Likewise, \emph{cost} is commonly measured by money value as well as effort in SPS studies for various purposes, such as \textit{coders cost}, \textit{estimated budget}, \textit{cost to repair defects} and so forth. 


\subsubsection{Resource metrics}
\begin{figure}[ht]
\centering
 \includegraphics[width=.5\textwidth]{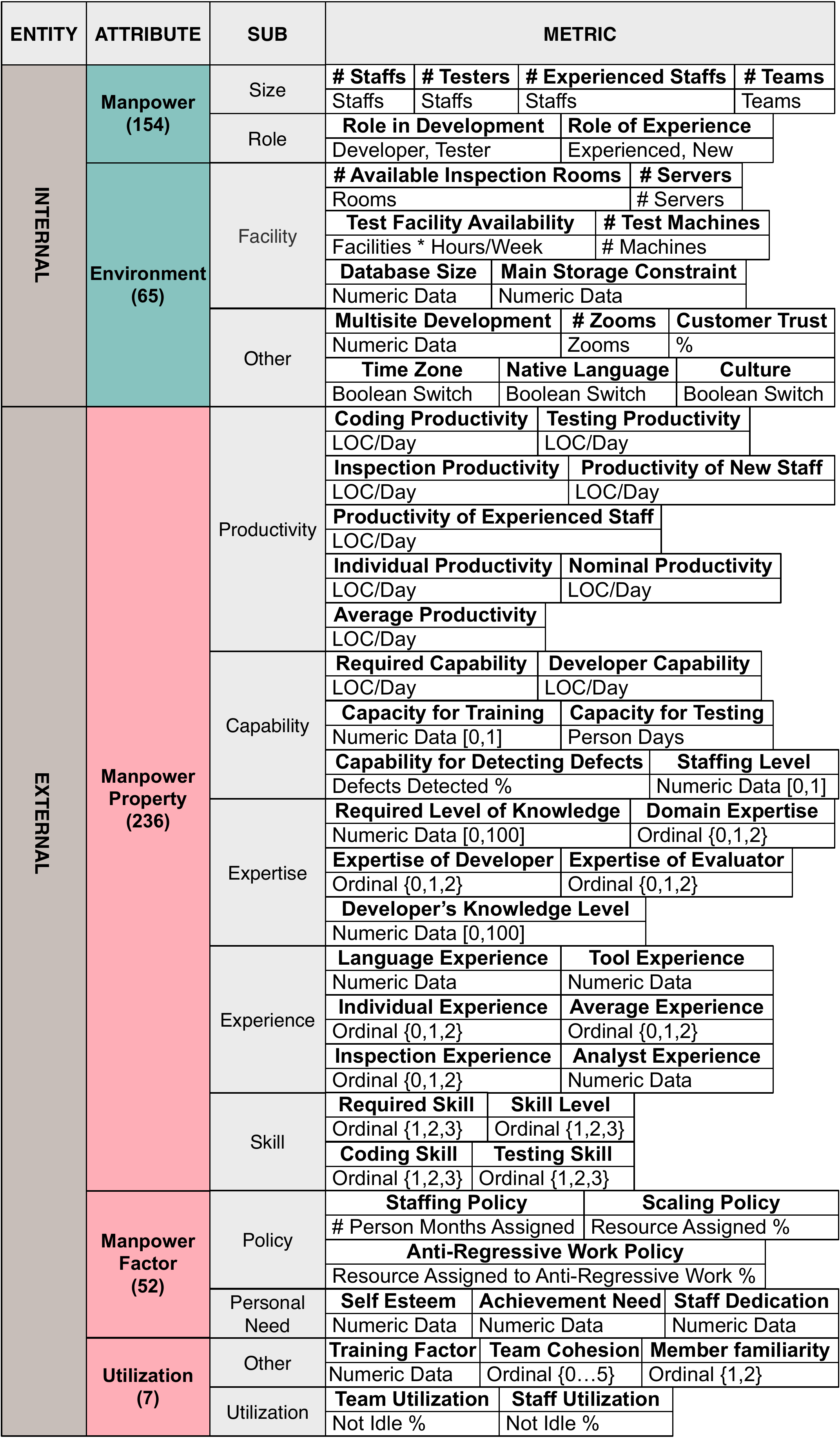}
\caption{Classification of Resource Metrics} \label{fig:metrics-resource}
\end{figure}

Different from product and process, \figurename~\ref{fig:metrics-resource} shows that modelers pay more attention to external entity for resource attributes since only the size or role of the resource can be measured by itself. The resource properties such as manpower skill cannot be measured only directly by itself without any other external reference. 

\textbf{\emph{Manpower}} measures the number of human resources for development. According to the modeling granularity and the scale of the simulated project, studies may regard all the resource as a team or allocate individuals to different phases. 

\textbf{\emph{Environment}} Few studies considered \emph{environment} in SPS models. As an example, \textit{test facility availability} is introduced to model the constraint of resource on test phase~\citesr{HoustonL10}. whether the project is \textit{multi-site development}, \textit{Multi-site development} is a factor suggested by COCOMO~\uppercase\expandafter{\romannumeral2}. 

To quantify the effect or the value of the manpower in a process, five basic \textbf{\emph{manpower property}}, i.e., \textit{productivity}, \textit{capability}, \textit{expertise}, \textit{experience}, and \textit{skill} can be measured. \textit{Productivity} forms the basis to contribute to the \textit{development rate}. The rest four metrics are the factors that determine \textit{productivity}. The rest four metrics are the factors that determine \textit{productivity}. The \textit{expertise} and \textit{knowledge} are the same metric. Different from \textit{expertise}, \emph{experience} already takes the domain knowledge into account, and were often simply measured by years without historical data.

\textbf{\emph{Manpower factor}} consists of all the influencing factors as metrics related to resource with a great diversity, ranging from \textit{self esteem}, \textit{team cohesion}, to \textit{native language}.

Only the human resource \textbf{\emph{utilization}} is considered in existing studies, although both individual and team \emph{utilization}s were used, they were only occurred in a few studies.

\begin{mdframed}[
	skipabove=.5\baselineskip
	innertopmargin=1ex,
	innerbottommargin=1ex,
	innerrightmargin=1ex,
	innerleftmargin=1ex,
	]
\textbf{Findings:} 1) For most reviewed studies only the metrics that are deemed to be significant are described in detail. 2) Product and process external metrics are not used frequently in process simulation modeling whilst resource external metrics are widely used.
\end{mdframed}

\subsection{Causal Relationships between Metrics (RQ2)}

The SPS models are built to simulate the process of the development team producing software products by carrying out a series of development-related activities under the constraints and support of the environment. In the process, activities will be organized together in a certain workflow and affect each other, and people are the specific actors of activities. People will be influenced by the environment and various other factors, and may also react to them. SPS models use different blocks (defined by paradigm) to depict different elements in the process, including people, activities, environments, etc. These blocks are instantiated implementations of the metrics discussed in this study in a specific simulation paradigm. At the implementation level, a SPS model is made up of blocks and their relationships to each other. The implementation of the model depicts different metrics and their interrelationships in reality. We discussed the metrics used in existing research in RQ1, and RQ2 will discuss causal relationships between metrics.

We collected all the causal relationships (from metric A to metric B) can be identified in included papers. As a result, we identified 183 types of causal relationships. Table~\ref{tab:casualrelation} presents high-frequency (with more than 10 occurrences in all SPS models and a relationship may appear multiple times in a model.) casual relationships and we discuss them in more detail below. 

From \textbf{\emph{manpower property}} to \textbf{\emph{manpower property}}. It is the most used causal relationship. The most direct relationship between software developers and software development activities is the metric \textit{productivity}, which describes a person's ability to participate in activities. Productivity is the most common one among the metrics of manpower property. \textit{Productivity} can be affected by factors such as \textit{experience} and \textit{knowledge}, which also belong to manpower property~\citesr{HanakawaMM98,klunder2018helping,van2018under}. In addition, \textit{productivity} can be more refined. For example, \textit{real-time productivity} can be composed of \textit{growth} and \textit{baseline/average productivity}~\citesr{Lehman2010Behavioural,Zawedde2013Determinants}. Metrics such as \textit{experience} can also be more refined. For example, the \textit{experience of inspectors} can be affected by \textit{development experience}, \textit{inspection experience}, and \textit{domain experience}~\citesr{neu2002simulation}.

From \textbf{\emph{defect size}} to \textbf{\emph{defect size}}. Defects can be in different states in the software life cycle. The change of state will be reflected in the change of quantity. The \textit{total of defects} can be the sum of \textit{open defects}, \textit{in progress defects}, \textit{waiting to test defects}, \textit{reopen defects}, and \textit{resolved and closed defects}~\citesr{zhang2018change}. Besides, The \textit{number of detected defects} can be affected by the \textit{number of generated defects}~\citesr{ZhangJZ08}. The \textit{number of escaped defects} can be the difference between the \textit{number of detected defects} and the \textit{number of fixed defects}~\citesr{Lak03}. Furthermore, defects can also be described as different types. The \textit{total of detected defects} can be the sum of \textit{detected passive defects} and \textit{detected active defects}~\citesr{ZhangKKJ08}.

From \textbf{\emph{change}} to \textbf{\emph{change}}. This type of causal relationships was mainly found in study~\citesr{zhang2018change} and study~\citesr{klunder2018helping}. In study~\citesr{klunder2018helping}, the \textit{sprint change capacity} is based on the \textit{daily change capacity}. The rest of relationships were found in study~\citesr{zhang2018change}, and the \emph{change} is also known as issue request in this work. The \emph{change} is modeled in detail based on the life-cycle of issues in the Issue Tracking System. For example, the \textit{number of issues waiting for review} is affected by the \textit{number of sprint issues}. The \textit{total number of issues} is the sum of \textit{duplicated and invalid issues}, \textit{open issues}, \textit{n progress issues}, \textit{sprint issues}, \textit{issues waiting for review}, \textit{resolved and closed issues}, and \textit{reopen issues}.

From \textbf{\emph{work rate}} to \textbf{\emph{work rate}}. One type of \textit{work rate} may be a composite of many other rates. For example, the \textit{nominal development rate} can be the sum of \textit{experienced employee development rate} and \textit{new employee development rate}~\citesr{Zhang2006Semi}. Besides, any type of \textit{actual work rate} can be based on a type of \textit{baseline work rate}~\cite{zhang2018change}.

From \textbf{\emph{artefact size}} to \textbf{\emph{artefact size}}. Artefact can have different origins or be in different stages of development. The \textit{project size} can be the sum of \textit{remaining size} and \textit{completed size}~\citesr{Zhang2006Semi}. The \textit{rate of generating requirements} can be affected by feedback of the \textit{number of generated new requirements}~\citesr{Zhang2009Qualitative}. The \textit{number of generated new requirements} can be based on the \textit{number of exogenous requirements}~\citesr{Hall2005Program}.  

From \textbf{\emph{artefact property}} to \textbf{\emph{artefact property}}. The refinement of \emph{artefact property} may not be so common, but it may be very fine-grained. Study~\cite{Alemran2008Simulating} used 15 kinds of different \textit{complexity} metrics to quantify the \textit{comprehensive complexity} of software systems. Study~\citesr{Aranha2008Using} modeled the relationships among \textit{number of control flows identified in the skeleton}, \textit{number of control flows written in the use cases that required rework after inspection}, and \textit{number of control flows written in the use cases}. Study~\citesr{Akerele2017} used \textit{ambiguity and brittleness of test code} to quantify the \textit{smell of test code}. 

From \textbf{\emph{process factor}} to \textbf{\emph{process factor}}. One \textit{process factor} can be broken down into a number of different factors. For example, the \textit{relevance of factoring} is based on a combination of multiple factors including \textit{relevance produktivity degree}, \textit{relevance of sprint course}, \textit{revlevance of reviews}, \textit{relevance of commitment-loyalty}, and \textit{relevance of customer satisfaction}.

From \textbf{\emph{time}} to \textbf{\emph{time}}. \emph{Time} usually describes the duration or delay of an activity. One activity contain a start time and an end time, and the difference between the two is the duration~\citesr{Zhou2012A}. The \textit{duration or delay} of an activity can be based on a \textit{mean or baseline value}~\citesr{Ruiz2002Integrating}. The \textit{average productive time} may be affected by the \textit{time loss due to work partitioning}~\citesr{HsiaHK99}.

From \textbf{\emph{manpower}} to \textbf{\emph{manpower}}. Manpower can be differentiated based on experience or other factors. The \textit{total number of employees} can be the sum of \textit{number of trained employees} and \textit{number of experienced employees}~\citesr{Zawedde2013Determinants}. The \textit{experienced employees} can be the sum of \textit{experienced employees for development} and \textit{experienced employees for training}~\textit{Zhang2006Semi}. The \textit{daily available workforce} can be based on the \textit{total workforce}~\citesr{Ambr2011Modeling}.

From \textbf{\emph{work rate}} to \textbf{\emph{artefact size}}. This type of relationship describes a common situation where the size of an aretefact accumulates based on the rate of work. For example, the \textit{specification units to be processed} is based on the \textit{specification unit completion rate}~\citesr{Wernick1999Software}. In addition, combining the \textit{software development rate} with other factors can also decide that \textit{requirements not met correctly}. Under this type of relationships, the latter is not a simple accumulation of the former~\cite{Hall2005Program}.

From \textbf{\emph{effort}} to \textbf{\emph{effort}}. Effort can be differentiated based on different activities. For example, the integration effort can be the sum of \textit{project assessment effort}, \textit{project tailoring effort}, and \textit{glue code effort}~\citesr{Naunchan2007Adjustable}. Furthermore, the \textit{development effort} can be calculated by the \textit{expected development effort} and \textit{growth effort}~\citesr{ChoiBK06}.

From \textbf{\emph{environment}} to \textbf{\emph{manpower property}}. The environment is the main factor that affects manpower. For example, the environmental factors such as \textit{market salary}, \textit{working environment}, \textit{team management} and \textit{reward} may affect the \textit{motivation} of developers~\citesr{fatema2018using}.

From \textbf{\emph{manpower property}} to \textbf{\emph{work rate}}. The rate of development (\emph{work rate}) is based on human productivity (\emph{manpower property}). For example, the \textit{new employee development rate} is based on the \textit{new hired workforce productivity}~\citesr{Zhang2006Semi}.

\begin{mdframed}[
	skipabove=.5\baselineskip
	innertopmargin=1ex,
	innerbottommargin=1ex,
	innerrightmargin=1ex,
	innerleftmargin=1ex,
	]
\textbf{Findings:} 1) High-frequency causal relationships are mainly relationships between metrics of the same type. It indicates that SPS models typically refine the main types of metrics such as \emph{manpower property}, \emph{defect size}, \emph{work rate}, \emph{change}, \emph{artefact size}, \emph{artefact property}, \emph{process factor}, \emph{time}, \emph{manpower}, etc. 2) Furthermore, even though the main line of the software life cycle is so clear, we do not see a relationship appearing in more than 10\% of the models. This fully demonstrates the diversity of SPS models, which requires specific analysis of specific problems as well as even the same problem may be implemented differently.
\end{mdframed}

\begin{table*}[htbp]
  \caption{High-frequency casual relationships between metrics}
    \begin{tabular}{p{30mm}p{30mm}p{15mm}p{90mm}}
    \toprule
    \multicolumn{2}{c}{\textbf{Casual Relationships}} & \multicolumn{1}{r}{\multirow{2}[4]{*}{\textbf{Frequency}}} & \multicolumn{1}{c}{\multirow{2}[4]{*}{\textbf{Citations}}} \\
    \cmidrule{1-2}
    \textbf{From}  & \textbf{To}    &       &  \\
    \midrule
    Manpower Property & Manpower Property & 44    & \citesr{Madachy2007Assessing,Lehman2010Behavioural,Zawedde2013Determinants,neu2002simulation,Neu2002Learning,ZhangJZ08,Cherif2010Software,Spasic2012Agent,van2018under,HanakawaMM98,fatema2018using,klunder2018helping} \\
    Defect Size & Defect Size & 22    & \citesr{Lak03,martin2000model,ZhangJZ08,ZhangKKJ08,zhang2018change} \\
    Change & Change & 22    & \citesr{zhang2018change,klunder2018helping} \\
    Work Rate & Work Rate & 20    & \citesr{Zhang2006Semi,WernickH02,van2018under,Wernick1999Software,zhang2018change} \\
    Artefact Size & Artefact Size & 19    & \citesr{Zhang2006Semi,Hall2005Program,Aranha2008Using,Zhang2009Qualitative,Wernick1999Software,zhang2018change,klunder2018helping,dugarte2021using} \\
    Artefact Property & Artefact Property & 18    & \citesr{Aranha2008Using,Shukla2012Dynamic,Akerele2017} \\
    Process Factor & Process Factor & 15    & \citesr{van2018under,zhang2018change,klunder2018helping} \\
    Time  & Time  & 15    & \citesr{HsiaHK99,Zhou2012A,Mishra2016A,WakelandMR04,Ruiz2002Integrating,zhang2018change,klunder2018helping} \\
    Manpower & Manpower & 12    & \citesr{Zawedde2013Determinants,Madachy2005Integrated,Zhang2006Semi,Ambr2011Modeling,Trammell2016Effects,Kahen2001System,zhang2018change} \\
    Work Rate & Artefact Size & 12    & \citesr{Lehman2010Behavioural,WernickH02,Hall2005Program,Zhang2009Qualitative,Wernick1999Software,zhang2018change,dugarte2021using} \\
    Effort & Effort & 10    & \citesr{ChoiBK06,Naunchan2007Adjustable,Aranha2008Using,noll2016teaching} \\
    Environment & Manpower Property & 10    & \citesr{HurtadoROT15,fatema2018using,klunder2018helping} \\
    Manpower Property & Work Rate & 10    & \citesr{Lehman2010Behavioural,Zhang2006Semi,MadachyT00,WernickH02,ZhangJZ08,Cocco2011Simulating,Mishra2016A,van2018under} \\
    \bottomrule
    \end{tabular}%
  \label{tab:casualrelation}%
\end{table*}%

\subsection{Metrics Selection Based on Purposes and Paradigms (RQ3)}

\begin{figure*}[!t]
\centering
  \includegraphics[width=.95\textwidth]{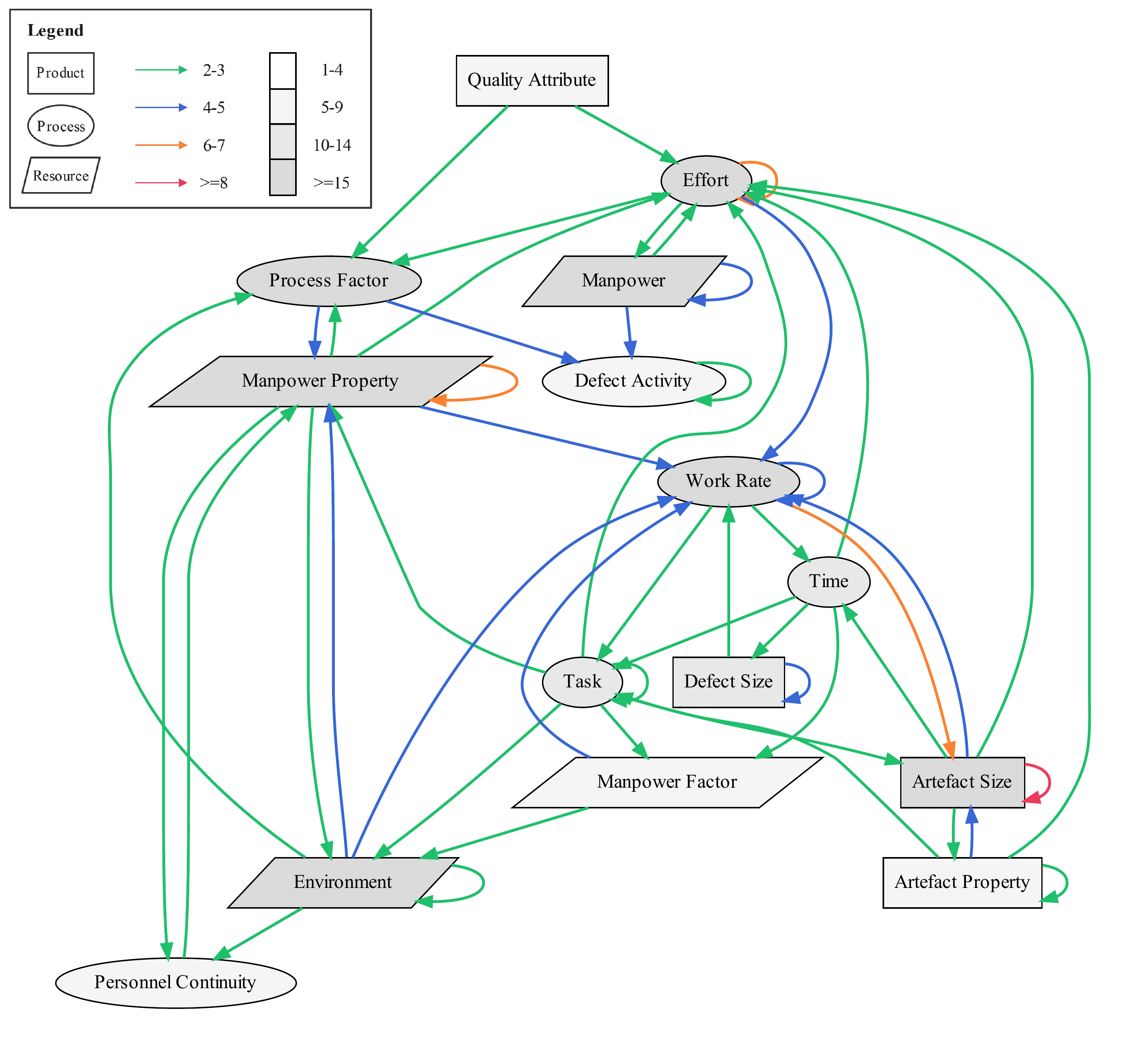}
\caption{Diagram of causal relationships between metrics in cognitive models (frequency $\geq 2$).}
\label{fig:casualrelation_l1}
\end{figure*}

\begin{figure*}[!t]
\centering
  \includegraphics[width=.8\textwidth]{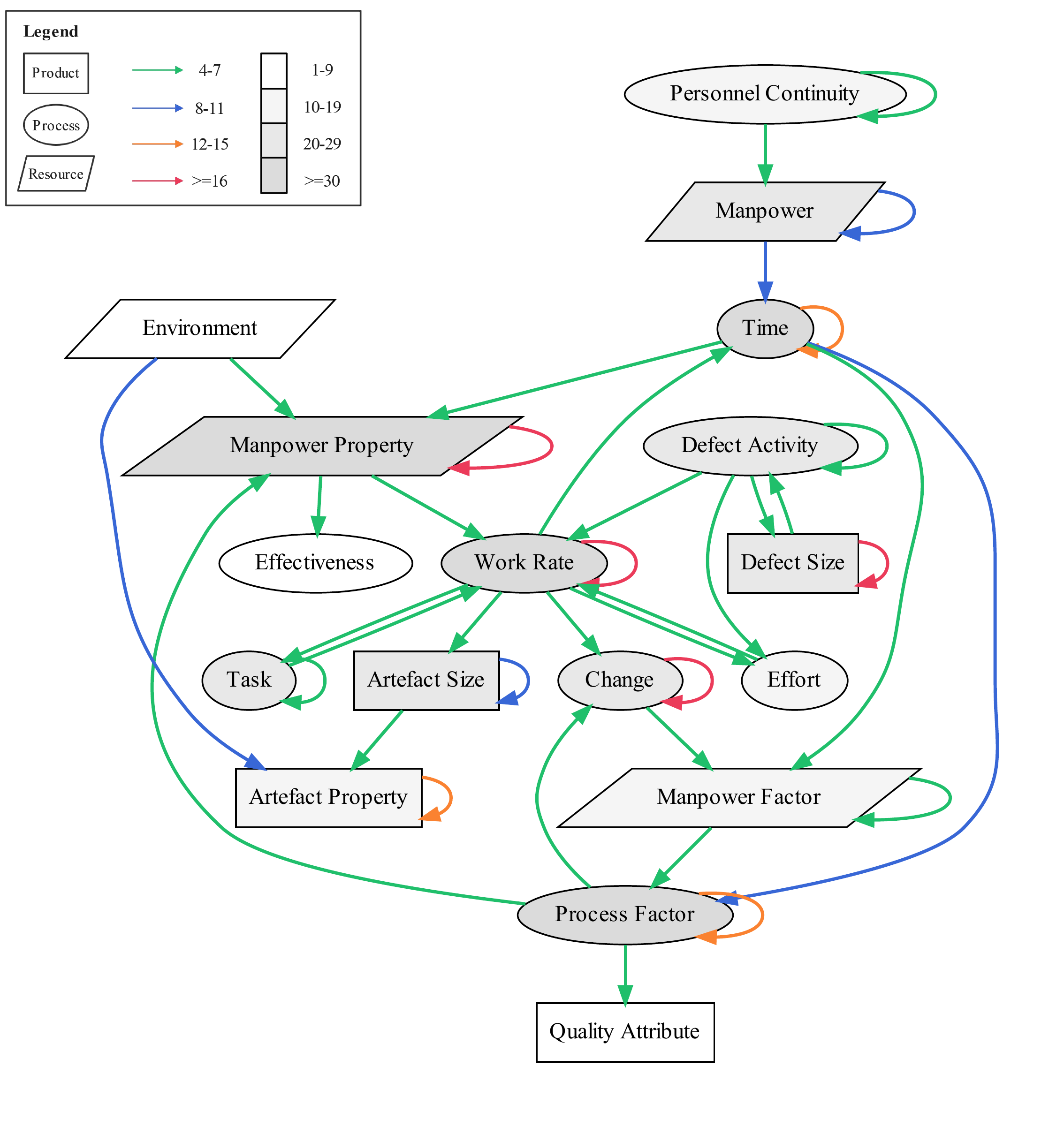}
\caption{Diagram of causal relationships between metrics in tactical \& strategic models (frequency $\geq 4$).}
\label{fig:casualrelation_l2}
\end{figure*}

As shown in the meta-model, the metrics selected for building the simulation model is determined by the information needs for building the corresponding conceptual model which depends on the modeling purposes. Kellner et al.~\cite{Kellner1999Software} also indicated that many aspects of what to simulate are inter-related and driven based on the purpose. Modelers simulate process at different granularity levels since it is nearly impossible as well as not necessary in some cases for a modeler to simulate every detail of a process. Hence, the selection of the appropriate metrics is critical to meet the needs of different modeling granularity. To a great extent, it depends on the specific purpose of a study. Existing SPS models were classified into Cognitive Models (CMs) and Tactical \& Strategic Models (TSMs) according to their modeling purposes as shown in \figurename~\ref{fig:purpose}. RQ3 discussed the commonalities and differences between CMs and TSMs from the perspective of causal relationships used.

In addition, the diversity and complexity of software processes, which can be reflected in modeling purposes, determine the different capabilities of simulation paradigms needed~\cite{ZhangKP10}. RQ3 will also discuss the differences in modeling granularity between different paradigms.

\subsubsection{Comparison between two levels of purposes in modeling}

We identified 217 causal relationships (the same type can be counted multiple times according to the frequency) in CMs and 462 causal relationships in TSMs. \figurename~\ref{fig:casualrelation_l1} and \figurename~\ref{fig:casualrelation_l2} present diagrams of causal relationships between metrics in CMs and TSMs respectively. If we included all the identified relationships, the diagram would be too complex to be understood. By compromise, we end up with a frequency greater than or equal to 2 times (about 1\% of the total) as a threshold for CMs. Considering that the number of relationships identified in TSMs is approximately twice as many as those identified in CMs. We set the threshold as 4 times for TSMs. As a result, the \figurename~\ref{fig:casualrelation_l1} contains a total of 51 types of causal relationships (the same type is only counted 1 time) consisting of 15 types of metrics, and the \figurename~\ref{fig:casualrelation_l2} contains a total of 39 types of causal relationships (the same type is only counted 1 time) consisting of 17 types of metrics.

In diagrams, we use box, oval, and parallelogram to denote product, process, and resource metrics respectively. The darker the fill indicates the higher the frequency of the metric used. We used different colored arrow curves to indicate the frequency of the relationship. The green, blue, orange, and red colors indicate frequencies greater than or equal to 2 (4), 4 (8), and 8 (16) for CMs (TSMs) respectively.

\textbf{From the perspective of metrics:} There is a clear overlap in the metrics of high frequencies in CMs and TMs, including \emph{artefact size}, \emph{defect size}, \emph{work rate}, \emph{task}, \emph{time}, \emph{process factor}, \emph{manpower}, and \emph{manpower property}. These all represent the basic elements of a software process, namely actors, artefacts, and activities. With these metrics, it is possible to describe a software development process at a macro level. This is probably the reason for their high frequency. The frequency of \emph{effort} is not high in TSMs but high in CMs. The frequencies of \emph{change} and \emph{defect activity} are high in TSMs but not high in CMs. Furthermore, there is \emph{effectiveness} in TSMs but not in CMs.

\textbf{From the perspective of causal relationships:} There are four types of relationships that are high-frequency in both CMs and TMs, including \emph{from artefact size to artefact size}, \emph{from defect size to defect size}, and \emph{from manpower to manpower}, \emph{from manpower property to manpower property}. It shows that CMs and TMs have commonalities in the fineness of modeling artefact, manpower, and defect, that is, to distinguish different artefact, manpower and defect, as well as the properties of manpower would be further refined. It should be noted that although \emph{from manpower property to manpower property} is a high-frequency relationship in both CMs and TSMs, its frequency is as high as 38 (8.2\%) in TSMs, but the frequency in CMs is only 6 (2.8\%). Besides, the frequencies of \emph{from defect size to defect size} in TSMs and CMs are 18 (3.9\%) and 4 (1.8\%) respectively. Most of the high-frequency relationships in CMs are in the form of a relationship \emph{from one metric to work rate}, whilst these are not high-frequency relationships in TSMs (except \emph{from work rate to work rate}). Even so, the types of relationships around \emph{work rate} are the most diverse in both CMs and TSMs. \emph{From time to process factor} and \emph{from manpower to time} are high-frequency relationships in TSMs, whilst there is no such relationships in CMs. \emph{From effort to effort} is high-frequency relationship in CMs, but not in TSMs. \emph{From artefact property to artefact property}, \emph{from time to time}, \emph{from process factor to process factor} are high-frequency relationships in TSMs, but not in CMs.

\begin{mdframed}[
	skipabove=.5\baselineskip
	innertopmargin=1ex,
	innerbottommargin=1ex,
	innerrightmargin=1ex,
	innerleftmargin=1ex,
	]
\textbf{Findings:} There are significantly more causal relations between metrics of the same type in TSMs, whilst relationships around \emph{work rate}, \emph{artefact size}, and \emph{effort} are more frequent in CMs. It indicates that CMs tend to directly model around the final result (e.g., effort, output artefact) at a macro-level granularity, whilst TSMs are more concerned with a micro-level granularity. TSMs tend to use multiple metrics within the same type to model an aspect (such as \emph{process factor}, \emph{manpower property}, \emph{defect size}, etc.) in detail.
\end{mdframed}

\subsubsection{Comparison of different modeling paradigms in modeling}

In order to simulate processes with different granularity, it is need to adopt appropriate modeling paradigms. For example, ABS is able to simulate each developer as an agent to study the impact of \textit{module complexity} on \textit{individual productivity} and \textit{motivation}. At this fine-grained level, where different developers need to be simulated separately, the model would be very complex using the SD paradigm. In the existing research, the DES and ABS paradigms have never been used in models that only stay at the cognitive level. Using different simulation paradigms mean that some of the metrics used will also change. Metrics that measure the same attribute need to change for different simulation paradigms, which is indirectly determined by the modeling granularity. Below are some examples to illustrate the main differences.

Probability metrics which indicates the probability that an event or state may change only used in DES and statistical models. For example, a DES model used \textit{the base probability $P(D_k^i(t))$} that specifies how likely it is that team $i$ will finish component $k$ after having worked on it for $t$ time units~\cite{padberg2002discrete}. If it is in an SD model, \textit{\# components per day} may be used to indicate the work rate. For paradigms that are good at macro-processes, like SD, it would be complicated to simulate activities that different teams implement different components into separate events.

For similar reasons, ABS models are better at modeling different developers in details. For example, Cherif and Davidsson~\cite{CherifD09} built an ABS model that simulates the difficulty of task $j$ as \textit{the difference between level of knowledge} $b_{ij}$ \textit{and the required level of knowledge} $\theta_j$, where $b_{ij}$ denotes \textit{developer} $i$ \textit{'s knowledge about activity} $j$. Cherif and Davidsson compared the ABS model with SD model and indicated that the main difference is that an SD model inputs average value of individual characteristics (\textit{manpower property}) as an alternative.

\begin{mdframed}[
	skipabove=.5\baselineskip
	innertopmargin=1ex,
	innerbottommargin=1ex,
	innerrightmargin=1ex,
	innerleftmargin=1ex,
	]
\textbf{Findings:} Compared with the SD model, DES and ABS can provide more detailed simulations from the perspective of development activities and individual developers, respectively, which is reflected in the more detailed metrics they use. In contrast, SD tends to use the mean as an alternative. This reduces the difficulty of obtaining metrics under the premise of less impact on the macro-process research.
\end{mdframed}



\subsection{RQ4: Revisit Data Issues and Coping Strategies}

As early as two decades ago, Raffo and Kellner et al.~\cite{Kellner1999Software,RaffoK00} discussed several situations that might arise when measuring process metrics in simulations and the possible coping strategies, which laid the foundation for subsequent research on SPS models. SPS models have encountered many new challenges over the years. Data issues were undoubtedly one of the major challenges of SPS.

As a result of reviewing 145 included papers, despite the fact that not every paper fully discussed their own data problems through modeling, we still summarized most of the data problems encountered during the data preparation stage in simulation modeling through retrieval. Furthermore, we included two related studies~\cite{Nauman2012,RaffoK00} as a supplementary data source for RQ4. We identified these two studies in the literature selection stage. Although these two studies did not meet our selection criteria C1 and C2, they reported experiences related to data issues. The inclusion of these two studies can help us to discuss data issues more comprehensively.

\begin{table*}[ht]
\centering
\scriptsize
\caption{Data issues encountered in SPS data preparing process (evidence from our included studies and related work).} \label{tab:data issues}
\begin{tabular}{p{20mm}p{15mm}p{62mm}p{35mm}p{30mm}}  
\toprule
\textbf{Data Preparing Steps} & \textbf{Encountered Data Issues}  & \textbf{Specific Issues} & \textbf{Metrics(E.g.,)} & \textbf{Ref.} \\
\midrule

\multirow{10}{*}{\textbf{Data Provenance}} & \multirow{4}{*}{\textbf{Availability}} & No readily available meta data in real-world projects & \textit{Rework Effort} & \multicolumn{1}{c}{\multirow{4}{30mm}{
\citesr{RuizRT01,RuizRT02,neu2002simulation,Berling2003Adaptation,Ruiz2004Using,ChoiBK06,Chen2006A,Turnu2006Modeling,Park2007Deriving,TopicJK08,Ferreira2009Understanding,Dan2013Modeling,Baum2017Comparing},~\cite{RaffoK00} }} \\
 & & Variables in the model required data not available in the literature  & \textit{Requirements Volatility; Affected Job Size(\%)} & \multicolumn{1}{c}{}  \\
 & & Records missing & \textit{Not Specified} & \multicolumn{1}{c}{} \\
 & & No quantitative information available & \textit{How TDD decrease Defects Injected Numbers} & \multicolumn{1}{c}{} \\
 \cmidrule{2-5}
 & \multirow{2}{*}{\textbf{Scarcity}} & Sample used were limited to a few projects & \textit{Not Specified} & \multirow{2}{30mm}{\citesr{Turnu2006Modeling,Concas2013Simulation,HurtadoROT15},~\cite{RaffoK00}} \\ 
 & & Scarcity of available information on samples & \textit{\#Maintenance Requests} & \\
 \cmidrule{2-5}
 & \multirow{2}{15mm}{\textbf{Reliability of metadata}} & Systematic biases are likely to exist in the repository data & \textit{Task's State} & \multirow{2}{*}{\citesr{,Kouskouras2007A,Baum2017Comparing},\cite{Nauman2012}} \\
 & & Expert estimates are impeded by a lack of intuition for some parameters & \textit{Not Specified} & \multicolumn{1}{c}{} \\
 \cmidrule{2-5} 
 & \textbf{Diversity} & Data diversity due to different organizational structures and project environments & \textit{Not Specified} & \cite{Nauman2012} \\
 \cmidrule{2-5}
 & \textbf{Traceability} & Data ``silos'' between different artifacts and sources & \textit{Requirements; Defects} & \cite{Nauman2012} \\ 
 \midrule
 \multirow{8}{20mm}{\textbf{Data Collection \&Processing}} & \multirow{2}{*}{\textbf{Understanding}} & Misunderstanding of the real meaning of the data & \textit{Not Specified} & \multirow{2}{*}{\cite{Nauman2012}} \\
 & & Difficult to find accurate descriptions for data fields because of distributed in various data sources & \textit{Not Specified} & \\
 \cmidrule{2-5}
 & \textbf{Accessibility} & Inability of various interfaces & \textit{Not Specified} & \cite{Nauman2012} \\
 \cmidrule{2-5}
 & \textbf{Time-span} & Difficulty in choice of that time-span of historical data & \textit{Not Specified} & \cite{Nauman2012},\citesr{Concas2013Simulation} \\\cmidrule{2-5} & \multirow{3}{*}{\textbf{Processing}} & Time and effort consuming & \textit{Software Developer Time \& Project Costs and Effort} & \citesr{PFAHL1999135,RaffoHV00,Ferreira2009Understanding} \\
 & & Considerable variability, outliers and noise & \textit{Not Specified} & \citesr{Concas2013Simulation,lunesu2017using},~\cite{RaffoK00}\\
 \midrule
 & \textbf{Definition} & Loosely defined & \textit{Not Specified} & \cite{RaffoK00} \\
 \cmidrule{2-5}
 & \textbf{Granularity} & Simulation model is an abstraction of real-world process, the metrics are coarse measures than actual data records. & \textit{Not Specified} & \citesr{DAVID2003SUPPORTING,Alemran2008Simulating,Zhang2012Simulation} \\
 \cmidrule{2-5}
 & \textbf{Accuracy} & Many of model's parameters could be estimated only,the accuracy of the measurements is questionable & \textit{Not Specified} & \citesr{Baum2017Comparing},~\cite{RaffoK00} \\
 \cmidrule{2-5} 
 \multirow{9}{*}{\textbf{Data Measurement}} & \textbf{Completeness} & Not possible to measure all variables and their relevance accurately & \textit{Not Specified} & \citesr{ChoiBK06},\cite{Nauman2012} \\
 \cmidrule{2-5}
 & \textbf{Measurement Method} & Existing studies rely on the use of industrial data averages, expert estimates, or values acquired from analytical models & \textit{Not Specified} & \cite{RaffoK00} \\
 & \multirow{2}{*}{\textbf{Quantifiability}} & Unquantifiable variables  & \textit{Skill Levels, Manpower, Communication Overhead, Review Support, Process Maturity, and Tool Support values} & \citesr{ChoiBK06,Armbrust2003Using,Concas2013Simulation,lunesu2017using} \\
 & & Unquantifiable variables' relationships & \textit{Interactions among developers-\textgreater{}Maintenance efforts} & \citesr{ChoiBK06,XiaoOWL10,Concas2013Simulation} \\
 \cmidrule{2-5}
 & \textbf{Dynamic} & Do not know the empirical distribution of variables variations & \textit{Growth and changes of developers' behaviors; effort variations; developers' allocation} & \citesr{Baum2017Comparing,lunesu2021assessing} \\
 \cmidrule{2-5}
 & \textbf{Reliability of outcome} & Does the above measure provide reliable & \textit{Not Specified} & \citesr{ChoiBK06,klunder2018helping,lunesu2021assessing} \\ 
\bottomrule                                                                                                                                                                               
\end{tabular}
\end{table*}

\subsubsection{Data issues encountered in SPS data preparing}

Specifically, we focused on the data issues encountered in the construction of SPS models. To facilitate understanding, we divided the data issues into three steps that align with the data preparation process of SPS, namely data provenance issues, data collection and processing issues, and data measurement issues.

\noindent\textbf{Data Provenance Issues.}

\textbf{\emph{1) Availability.}} Software processes generate all sorts of complex data, and the availability of data has always been the greatest challenge in simulation modeling research. It was evident from all 17 studies that data availability was a problem. Not all activities are documented, and most of the data are not available through relevant literature. There are missing data records, for example, there are no records for \textit{rework efforts} on any activities \citesr{ChoiBK06}. In particular, quantitative data sources are unavailable (e.g., \textit{rate of job size added by requirements volatility}~\citesr{Ferreira2009Understanding}).

\textbf{\emph{2) Scarcity.}} Even if data are obtained, the data scarcity problem is a common issue for modelers. In most simulation studies, only a few data were available for simulation since the histories of selected projects are too short~\citesr{Concas2013Simulation}. A general problem with the record data in projects is that it \emph{does not provide adequate information} to simulate the real variable change and interrelationship~\citesr{Turnu2006Modeling}.

\textbf{\emph{3) Reliability of metadata.}} Modelers would be concerned about the authenticity and reliability of the data when they come into contact with the real data source. Both quantitative and qualitative data is obtained from software repositories, manual record documents and expert estimation whilst these data sources are more or less error-prone. For example, software developers may incorrectly choose defect types, forget the defect location time, delay in updating the requirements and planning schedule documents~\citesr{Baum2017Comparing},\cite{Nauman2012}.

\textbf{\emph{4) Diversity.}} As a result of the differences in data types, record forms, project processes, and organizational structures, there will naturally be a variety of process data \citesr{HurtadoROT15}. Because of different organizational and project development environments, data sources can be distributed in a variety of online and offline sources without direct correlation. If data from multiple stages and activities of the software process are involved in the simulation model, it is challenging to ensure the consistency of the data.

\textbf{\emph{5) Traceability.}} It is of interest to restore the true evolution of process data by building traceability between them. There is a lack of discussion of this aspect of the problem in the previous studies and how they deal with it. Ali et al. \cite{Nauman2012} noted that real-world projects have ``silos'' between process data information. The requirements repository stores process information about requirements, while defect reports are stored in another database, i.e. issue tracking system. As a result, despite the existence of both data points, we are unable to use them since their connections have been missed.

\noindent\textbf{Data Collection \& Processing Issues.}

Even the data sources for modeling are available, the modeler still struggles with how to collect the data and how to process it to meet subsequent simulation modeling demands.

\textbf{\emph{1) Understandability.}} The process data may distributed in various data sources across various departments, therefore it might be difficult to understand accurate meanings of descriptions for data-fields in documentation~\cite{Nauman2012}. Modelers must consult domain experts frequently about data fields, even if the name seems straightforward. The field names differ between organizations, teams, and databases. For instance, in the issue tracking system, the release version number may be defined as ``baseline'', whilst it might be defined as ``version'' in the requirements management repository.

\textbf{\emph{2) Accessibility.}} In order to export data from multiple data sources, diverse interfaces must be accessible. Accessing the interface still requires permissions and authorizations from various business departments, making the data collection process extremely complicated and time-consuming. A risk of data collection failure exists due to data sensitivity and permission concerns.

\textbf{\emph{3) Time-span.}} It has been recommended by Ali et al.~\cite{Nauman2012} to discuss the selection of historical data time-span for the construction and calibration of SPS models with domain experts. In order to simulate the change distribution of variables and the interaction between them, a long enough time period is required~\citesr{Concas2013Simulation}. When simulating the current reality with historical data, sufficient historical data is required so that we can understand potential changes adequately. For example, if the development process, programming language, or development platform changes, modelers need to be aware of these changes.

\textbf{\emph{4) Processing.}} Obtaining, collecting, processing and analyzing data is undoubtedly a complex process that requires considerable time and effort~\citesr{PFAHL1999135, Ferreira2009Understanding, RaffoHV00} due to issues related to data diversity, reliability, understandability, etc. The current data collection and processing activities in SPS research still relies on manual methods, e.g., data format conversion, outlier and noisy data processing. 

\noindent\textbf{Data Measurement Issues.}

The primary purpose of data preparation should be to provide the data necessary for the simulation model to measure model metrics, which requires measuring the real values of model variables and their relationships. Due to this, SPS places a great deal of importance on the measurement of metrics.
Based on a summary of measurement issues reported in previous works, we classified them into eight categories, which are definitions, granularity, accuracy, completeness, measurement method, quantiability, and dynamic of metrics, as well as the reliability of outcome of SPS models.

\textbf{\emph{1) Definition.}} The meta-model illustrates that the metrics used in SPS models are based on the information needs used in the construction of the conceptual model, which depends on the purpose of modeling. Due to cognition differences, modelers may not be able to comprehend the details of the development process. They might ignore the difficulty of measurement and define inappropriate metrics as a result.

\textbf{\emph{2) Granularity.}} Often a modeler cannot simulate every detail of a process because it is either impossible or not necessary to do so. Metrics should therefore be selected and measured based on the modeling granularity. However, in many cases, real-world data is too fine-grained to match the assumptions of the model~\citesr{Zhang2012Simulation}. As an example, SD model assumes no difference between the input variables (e.g., new requirements). However, there is a huge difference between different requirements in actual circumstances, including the difficulty, priority, dependency, and the amount of developers should be invested in them~\citesr{Alemran2008Simulating}. Both of these factors may affect the size of the job.

\textbf{\emph{3) Accuracy.}} The accuracy of measurement is also a common problem in SPS modeling. The accuracy is often questioned due to the quality of the data and the choice of measurement methods. When available data sources are limited, many variables in the model have to be estimated by experts or referred to other literature. Expert estimates, however, are derived from subjective perceptions and always hampered by a lack of intuition~\citesr{Baum2017Comparing,klunder2018helping}.

\textbf{\emph{4) Completeness.}} Moreover, not all variables can be accurately defined and measured. Due to issues relating to data availability, quality, and metric measurement, modelers have to adjust the structure and scope of the model, which may also affect the completeness of SPS models.

\textbf{\emph{5) Measurement methods.}}
Measurement methods typically include mean value estimation, probability estimation, distribution fitting, and stochastic simulation using Monte Carlo. Moreover, existing SPS studies use industry averages, expert empirical estimates, or values derived from analytical models \cite{Bernard2020}. These methods are far too simplistic and wild to meet the modeling demands of capturing real changes of variables and their reciprocal effects.

\textbf{\emph{6) Quantifiability.}} Many process metrics, such as human aspect factors \citesr{ChoiBK06} (e.g., skill levels, interactions among developers), are difficult to quantify, but are extremely critical for the simulation model because they influence productivity greatly \citesr{lunesu2017using}. Quantifying the causal-relationships between variables accurately is a classic challenging task\cite{Nauman2012}.

\textbf{\emph{7) Dynamic.}} Despite improvements in data sources and quality, modelers still are not able to fit the true dynamic distribution of variables from historical data. For example, it is difficult to measure the growth of developers as well as the impact of the growth on their behavior~\citesr{lunesu2021assessing}.

\textbf{\emph{8) Reliability of outcome.}} All of the data preparation work ultimately serves to construct and calibrate the simulation model. This in turn would affect the reliability of the simulation results directly. In other words, it is challenging to validate and ensure that the artefacts, activities, and actors modeled in an SPS model are reliable enough to be able to generate plausible results~\citesr{ChoiBK06,lunesu2021assessing}. This is something we ultimately need to address to meet the modeling purpose.

\begin{mdframed}[
	skipabove=.5\baselineskip
	innertopmargin=1ex,
	innerbottommargin=1ex,
	innerrightmargin=1ex,
	innerleftmargin=1ex,
	]
\textbf{Findings:} 
1) Throughout the evolution of modeling work, the availability of data has been a major concern.
2) Additionally, modeling requirements for the model metrics (i.e. modelling changes of real process variables and the mutual influence relationships) present a big challenge to modelers.
3) There would be specific problems encountered at various stages of the preparation process, including problems with reliability, diversity, traceability, historical data time span, the data interface, the format of the data, and the processing time, etc.
\end{mdframed}

\subsubsection{The relationship between data issues and coping strategies}

\begin{figure*}[ht]
\centering
 \includegraphics[width=0.8\textwidth]{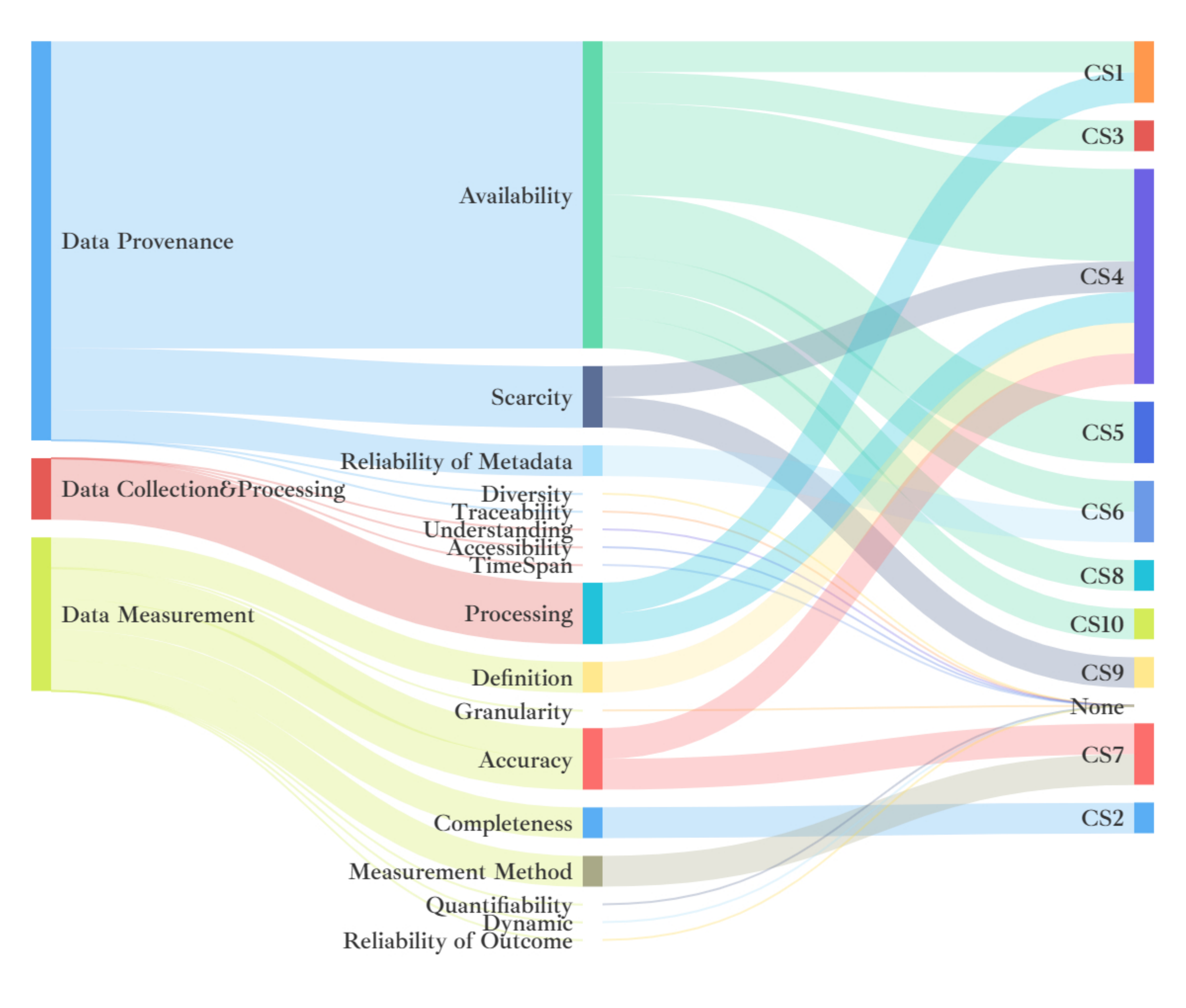}
\caption{The relationship between data issues and coping strategies} \label{fig:relationship}
\end{figure*}

\begin{table}
\centering
\scriptsize
\setlength\tabcolsep{4.5pt}
\caption{Coping strategies} \label{tab:coping strategies}
\begin{tabular}{p{3.5mm}p{50mm}p{25mm}}
\toprule
\textbf{ID} & \textbf{Coping Strategies}& \textbf{Ref.} \\ \hline
CS1         & Directly go to the source documents and repositories.            & \citesr{Baum2017Comparing,RaffoHV00,Menzies2002Model}        \\
CS2         & Look for data in other parts of the organization.                                                        & \citesr{Menzies2002Model},\cite{Nauman2012}                       \\
CS3         & Develop a survey to collect the needed data.                                                             & \citesr{Menzies2002Model,Ferreira2009Understanding}                                                         \\
CS4         & Work with experienced process participants or using experts' opinion, e.g., developer, project manager.  & \citesr{PFAHL1999135,Menzies2002Model,Chen2006A,TopicJK08,RusNM14},~\cite{RaffoK00} \\
CS5         & Get data from the literature.                                                                            & \citesr{Berling2003Adaptation,Ruiz2004Using,Turnu2006Modeling,RusNM14}                                     \\
CS6         & Use approximate data or easily configurable data.                                                        & \citesr{Kouskouras2007A}                                                             \\
CS7         & Adjust model variables: another calculated method, using replacement metrics, adjust scope of variables. & \citesr{Baum2017Comparing},~\cite{RaffoK00}                                                        \\
CS8         & Adjust model scope.                                                                                      & \cite{RaffoK00}                                                                                         \\
CS9         & Adjust the scale of historical data.                                                                     & \citesr{Concas2013Simulation}                                                                              \\
CS10        & Drop the variable from the model.                                                                        & \cite{RaffoK00} \\

\bottomrule                                                                                        
\end{tabular}
\end{table}

\begin{table*}[h]
\centering
\scriptsize
\caption{Available data sources} \label{tab:available data sources}
\begin{tabular}{p{60mm}p{70mm}p{35mm}}  
\toprule
\textbf{Data sources}&\textbf{Metric} & \textbf{Ref.} \\ 
\midrule
  \textbf{Experienced experts} \\ \hline
  Experienced programmers and testers & \textit{Amount of incoming work}; \textit{\# programmers}; \textit{\# injected defects per day per programmer}, etc.& \citesr{Berling2003Adaptation}\\
  Industrial project managers' intuitions and experience& \textit{workload for the activity}; \textit{developers' abilities}; \textit{development duration} & \citesr{Hanakawa2002A}\\
  A manager in the distributed support department & \textit{range of potential productivity improvements} & \citesr{Pfahl2004PL}\\
  Survey with professionals of different areas& \textit{effort} & \citesr{Shukla2012Dynamic}\\
  Past organizational data and expert judgment & \textit{estimated requirements volatility} & \citesr{MadachyT00}\\
  Expert judgment & \textit{review skill}; \textit{global blocker issue risk}; \textit{global blocker issue suspend time}; \textit{review remark fix duration}; \textit{issue assessment duration}; \textit{planning duration}; \textit{task switch overhead}; \textit{review fix to task factor}, etc.& \citesr{Baum2017Comparing}\\

\midrule
  \textbf{Literature} \\ \hline
  Abts et al.~\cite{Abts1997} & \textit{delivery time}; \textit{productivity}; \textit{cost} & \citesr{Ruiz2004Using}\\
  Lum et al.~\cite{Karen2002}& \textit{cost} & \citesr{Uzzafer2013A}\\
  Jones et al.~\cite{Jones1996Applied}& \textit{\# defects}; \textit{costs}; \textit{duration covered in the QA model}& \citesr{DrappaL99}\\
  Frost et al.~\cite{Frost2007Advancing}& \textit{defect injection rate}; \textit{verification effectiveness} & \citesr{Garousi2009A}\\
  Wagner et al.~\cite{Wagner2006LSQ}& \textit{typical (average) verification and validation rate}; \textit{rework effort for defects of various document types}& \citesr{Garousi2009A}\\
 \midrule

\textbf{Organization's documents} \\ \hline
  Project management documents & \textit{schedule}; \textit{effort}; \textit{code size} & \citesr{DIETMAR2000KNOWLEDGE,Menzies2002Model,RaffoHV02}\\
  Individual inspection reports& \textit{defect detection rate}; \textit{inspection effectiveness}& \citesr{Menzies2002Model,RaffoHV02}\\
  Holon history records& source code;change history data; estimate of WIP& \citesr{Chatters2000Modelling}\\
  Company's ticket system& \textit{defect injection rate}; \textit{task duration}; \textit{review duration}; \textit{issue fix overhead}; \textit{\# developers}; \textit{dependency} & \citesr{Baum2017Comparing}\\
  Time reporting system & \textit{ratio of new development to rework} & \citesr{Pfahl2004PL}\\
  Configuration management system & move-to-production statistics & \citesr{Pfahl2004PL}\\
  IT recruiting and training department & \textit{attrition rate}; training statistics& \citesr{Pfahl2004PL}\\
\bottomrule
\end{tabular}
\end{table*}

We identified ten coping strategies (CS1-CS10) as shown in Table~\ref{tab:coping strategies}, of which most of them have already been discussed by Raffo and Kellner~\cite{RaffoK00}, CS3 and CS9 are newly discovered strategies. 
Figure~\ref{fig:relationship} shows a visualization map that allows to assess the relationship between data issues, the strategies used to cope with them, and the number of evidence in SPS studies provided to support those conclusions. 

Since data availability has been a major challenge for modelers over the years, researchers have proposed a variety of solutions. When it is possible for a modeler to access multiple data sources, they could simultaneously use the source documents, repositories, and working with experienced practitioners, getting estimate values from experts, getting data from the literature, etc., to reduce bias. 
There three types of common data sources can be used when the available data for modeling is not sufficient as shown in Table~\ref{tab:available data sources}, which are experienced experts, literature, and organization's documents.

\textbf{\emph{Experienced experts (CS4)}} is one of the main data sources for collecting and verifying metrics. For example, a manager in the distributed support department was conducting a process improvement program and provided us with an estimate of the range of potential productivity improvements~\citesr{Pfahl2004PL}. Furthermore, advice from experts is especially needed for qualitative data.

\textbf{\emph{Literature (CS5).}} In some cases, If project data are not available statistical, metrics and typical functions from literature can be used initially. A significant part of the cause and effect relationships covered in the Quality Assurance model is quantitatively supported by the data provided in the study~\cite{Jones1996Applied}, and they provides data concerning errors, costs and duration. In the study~\cite{Abts1997}, the effect of different integration starting points on delivery time, productivity and cost were analyzed. To calibrate GENSIM 2.0, Frost et al.~\cite{Frost2007Advancing} provided an example of a defect containment matrix from which values for the calibration fault injection rates and verification effectiveness can be derived. Wagner~\cite{Wagner2006LSQ} provided much data on typical (average) verification and validation rates, and rework efforts for defects of various document types.

\textbf{\emph{Organization's documents (CS1, CS2).}} Organizations maintain a wide variety of documents which contains a number of metrics. Different organizations adopt different processes and use different tools, hence, the available evidence could not cover all types of documents.

If measurement data sources are not available, it is suggested to use an approximation value (CS6) or adjust variables' measurement methods (CS7). In case of all the above measures fail, modelers may need to simplify the scope and structure of the model again, or maybe just drop the variables.

Studies\citesr{PFAHL1999135,Menzies2002Model,Chen2006A,TopicJK08,RusNM14},~\cite{RaffoK00} indicated that modelers work closely with experienced practitioners to fully understand software processes, so as to build models based on an in-depth understanding of the data. 
It is necessary to consult internal experts to determine whether the project has historical data of a certain scale within it and whether the necessary information has been recorded so as to avoid the issue of data scarcity in modeling.

As a means of avoiding repetitive data collection and processing steps due to information asymmetry and cognitive differences, researchers recommend integrating software repositories, data documents, and internal practitioners' opinions together. Without those strategies, manpower and time would be wasted. There is little discussion of misunderstandings, data accessibility risks, or time spans for history data, even though these issues were raised by Ali et al.~\cite{Nauman2012} and other modelers may face similar challenges.


There are not many novel strategies provided by researchers for data measurement. Aside from that, we observed that the measurement problem still has a number of difficulties, such as matching modelling granularity, quantifying variable models, fitting dynamic distributions to variables, etc., for which there were few papers that gave appropriate countermeasures. Undoubtedly, this will result in modeling bottlenecks. As a result of this situation, the subsequent simulation work might be held to an even higher standard. A subsequent study is expected to examine these problems in more detail and provide comprehensive and detailed solutions, thereby serving as a useful guide and practice for modelers.

\begin{mdframed}[
	skipabove=.5\baselineskip
	innertopmargin=1ex,
	innerbottommargin=1ex,
	innerrightmargin=1ex,
	innerleftmargin=1ex,
	]
\textbf{Findings:} 1) We found that relatively few SPS studies discussed appropriate strategies for coping with specific data quality problems from the data provenance aspect, such as reliability of metadata, diversity of process data, and data traceability, etc. 
2) There are also a lack of details regarding effective communication with practitioners, minimizing data accessibility risks, and selecting the appropriate time-span for project history information.
3) In terms of data measurement, researchers have not provided many novel strategies. However, it is really critical to solve these issues in the construction and application of SPS models in real-life projects. Further studies are expected to provide comprehensive and detailed solutions to these issues.
\end{mdframed}

\section{Threats to Validity} 
We check the validity from four aspects based on the mapping provided by Zhou et al.~\cite{ZhouJZLH16}.

\textbf{Construct Validity:}
Finding all relevant papers was a common threat for all SLR studies.  We may have missed a few related studies for SPS in paper selection process
To overcome this, we have done extensive research on SPS and using the search string has been validated in previous studies. To search the papers published after 2015, we also used Wohlin's forward snowballing strategy~\cite{Wohlin16}.

\textbf{Internal Validity:}
Selection bias is also a standard threat to all SLRs. Our selection method was based on the QGS method~\cite{ZhangBT11} and snowballing in order to reduce bias. A forward snowballing strategy was employed using three seminal papers which were cited by most relevant studies with higher citations in SPS research in order to extend the pool of relevant studies until 2021.
Another threat was how to minimize inaccuracies when extracting data. Study selection and data extraction are mainly carried out independently by two students, and they conduct a cross check before synthesizing the data.
The supervisor and students met weekly to discuss all the inconsistencies and disagreements.
One threat is that most studies did not provide a complete list of the metrics they used, much less present the details of whether a metric is calibrated or not. We only extracted the metrics and their relationships reported in the study that were considered as the important and general metrics by the authors.

\textbf{External Validity}: We ensure that the primary study of the selection is high generalizability which may lead to the high generalizability of the study conclusion.

\textbf{Conclusion Validity:}
During the data synthesis phase of the SLR, grouping extracted metrics and data issues was another challenge. Based on Fenton et al.'s classification~\cite{Fenton2014Software} on metrics into three categories, we strictly followed the definitions of the different categories to ensure the correctness of metric classification. To construct a systematic set of findings, we applied the thematic synthesis method~\cite{Cruzes2011Recommended} and coding technique~\cite{Corbin1998Basics} in an iterative process. 
This allowed us to develop a consistent set of codes based on the diverse descriptions of metrics. By applying thematic synthesis~\cite{Cruzes2011Recommended}, we summarized the data issues encountered in the construction of SPS models by aligning them with the data preparation process of SPS.

There is also a common threat for us, which is the lack of details from primary studies. Most of our primary studies did not provide such details because they generally focused on building models and analyzing their results. Therefore, we were unable to extract and synthesize them comprehensively and completely. It is therefore expected that the subsequent study on SPS will examine more problems in more detail and provide comprehensive and detailed information regarding SPS metrics. This includes measurement methods, values, categories, data issues, coping strategies, and data sources, etc. This serves as a useful guide and practice for modelers.

\section{Conclusion}
SPS modelers bear the burden of the high cost of selecting appropriate metrics, 
which is regarded by the community as one of the crucial barriers in adopting SPS in practice. This paper reports an SLR on the metrics and the associated attributes used in SPS models. The implication of this work can be extended to software process modeling in general considering the continuity and similarity between the static models and dynamic models and the higher standard required for dynamic models. The study is driven by a meta-model of the ontology of metrics so that we are able to comprehensively study the metrics in SPS models from the perspective of modeling. 

As the result of an exhaustive search, we identified a total of 145 papers report SPS models until 2021. We identified 2130 metrics from these models and classified metrics and their corresponding attributes into six entity categories from the measurement perspective. Coding technique is applied to build the classification framework. The framework is more suitable as a reference for modeling than existing software measurement frameworks since it includes all metrics limited to SPS and has multiple levels of categories to show the similarities and differences of metrics. Different from Fenton et al.'s framework~\cite{Fenton2014Software}, we distinguish \textit{defect activity} from \textit{defect size} and \textit{defect property} as well as classify \textit{defect size} and \textit{defect property} into product category since it is more reasonable. The choice of paradigm is constrained by both modeling granularity and available data. Therefore we further discussed data issues of measurement and coping strategies.

This study helps to address one of the two major challenging tasks of modeling, i.e. identifying the key metrics (elements of the model) and their relationships (structure of the model) from real processes. Although we discussed data issues and coping strategies reported in existing studies, it is still a challenging task for modelers to collect the data from industrial settings. SPS studies usually assume that organizations can collect required information but it is recognized that this is not feasible in some cases. In recent years process mining turns to be an effective tool to distill process information from various software repositories. Many mining algorithms and tools aiming at different purposes and data formats are mapped in our previous study~\cite{Dong17MSP}. Investigating how to apply an appropriate mining technique to obtain the data on the required metrics can be an opportunity for future work to improve the practical adoption of SPS.

Furthermore, the metrics and knowledge synthesized from SPS studies are not limited to process simulation but can be extended to software process modeling in general. Taking simulation metrics as standards and references can further drive software developers to improve the collection, governance and application of process data in practice.

\ifCLASSOPTIONcompsoc
  \section*{Acknowledgments}
\else
  \section*{Acknowledgment}
\fi
This work is supported by the National Natural Science Foundation of China (No.62072227, No.62202219), the National Key Research and Development Program of China (No.2019YFE0105500) jointly with the Research Council of Norway (No.309494), the Key Research and Development Program of Jiangsu Province (No.BE2021002-2), the Intergovernmental Bilateral Innovation Project of Jiangsu Province (No.BZ2020017), as well as the Innovation Project and Overseas Open Project of State Key Laboratory for Novel Software Technology (Nanjing University) (ZZKT2022A25, KFKT2022A09).

\ifCLASSOPTIONcaptionsoff
  \newpage
\fi




\renewcommand{\refname}{References}
\bibliographystyle{IEEEtran}
\bibliography{spsmetrics2018}

\appendix
\section{References of the Reviewed Studies} \label{sec:appendix}
\renewcommand{\refname}{Bibliography}
\bibliographystylesr{wmaainf}
\bibliographysr{sr-list}

\end{document}